\documentstyle[psfig]{mn}

\newif\ifAMStwofonts

\ifoldfss
  \ifCUPmtlplainloaded \else
    \NewTextAlphabet{textbfit} {cmbxti10} {}
    \NewTextAlphabet{textbfss} {cmssbx10} {}
    \NewMathAlphabet{mathbfit} {cmbxti10} {} 
    \NewMathAlphabet{mathbfss} {cmssbx10} {} 
  \fi
  \ifAMStwofonts
    \ifCUPmtlplainloaded \else
      \NewSymbolFont{upmath} {eurm10}
      \NewSymbolFont{AMSa} {msam10}
      \NewMathSymbol{\upi}     {0}{upmath}{19}
      \NewMathSymbol{\umu}     {0}{upmath}{16}
      \NewMathSymbol{\upartial}{0}{upmath}{40}
      \NewMathSymbol{\leqslant}{3}{AMSa}{36}
      \NewMathSymbol{\geqslant}{3}{AMSa}{3E}

    \fi
  \fi
\fi 

\ifnfssone
  \newmathalphabet{\mathit}
  \addtoversion{normal}{\mathit}{cmr}{m}{it}
  \addtoversion{bold}{\mathit}{cmr}{bx}{it}
  \newmathalphabet{\mathbfit} 
  \addtoversion{normal}{\mathbfit}{cmr}{bx}{it}
  \addtoversion{bold}{\mathbfit}{cmr}{bx}{it}
  \newmathalphabet{\mathbfss} 
  \addtoversion{normal}{\mathbfss}{cmss}{bx}{n}
  \addtoversion{bold}{\mathbfss}{cmss}{bx}{n}
  \ifAMStwofonts
    \ifCUPmtlplainloaded \else
      \UseAMStwoboldmath
      \makeatletter
      \new@mathgroup\upmath@group
      \define@mathgroup\mv@normal\upmath@group{eur}{m}{n}
      \define@mathgroup\mv@bold\upmath@group{eur}{b}{n}
      \edef\UPM{\hexnumber\upmath@group}
      \new@mathgroup\amsa@group
      \define@mathgroup\mv@normal\amsa@group{msa}{m}{n}
      \define@mathgroup\mv@bold\amsa@group{msa}{m}{n}
      \edef\AMSa{\hexnumber\amsa@group}
      \makeatother
      \mathchardef\upi="0\UPM19
      \mathchardef\umu="0\UPM16
      \mathchardef\upartial="0\UPM40
      \mathchardef\leqslant="3\AMSa36
      \mathchardef\geqslant="3\AMSa3E
    \fi
  \fi
\fi 

\ifnfsstwo
  \DeclareMathAlphabet{\mathbfit}{OT1}{cmr}{bx}{it}
  \SetMathAlphabet\mathbfit{bold}{OT1}{cmr}{bx}{it}
  \DeclareMathAlphabet{\mathbfss}{OT1}{cmss}{bx}{n}
  \SetMathAlphabet\mathbfss{bold}{OT1}{cmss}{bx}{n}
  \ifAMStwofonts
    \ifCUPmtlplainloaded \else
      \DeclareSymbolFont{UPM}{U}{eur}{m}{n}
      \SetSymbolFont{UPM}{bold}{U}{eur}{b}{n}
      \DeclareSymbolFont{AMSa}{U}{msa}{m}{n}
      \DeclareMathSymbol{\upi}{0}{UPM}{"19}
      \DeclareMathSymbol{\umu}{0}{UPM}{"16}
      \DeclareMathSymbol{\upartial}{0}{UPM}{"40}
      \DeclareMathSymbol{\leqslant}{3}{AMSa}{"36}
      \DeclareMathSymbol{\geqslant}{3}{AMSa}{"3E}
    \fi
  \fi
\fi 

\ifCUPmtlplainloaded \else
  \ifAMStwofonts \else 
    \def\upi{\pi}
    \def\umu{\mu}
    \def\upartial{\partial}
  \fi
\fi

\title{Near infrared spectroscopy of starburst galaxies}
\author[R. Coziol, R. Doyon \& S. Demers]
       {R. Coziol,$^{1,2}$ R. Doyon,$^3$ S. Demers,$^3$\\
1- El Departamento de Astronomia de la Universidad de Guanajuato, 
Apartado Postal \#144, 36000 Guanajuato, Gto M\'exico\\
2- Observatoire de Besan\c{c}on, UPRES--A 6091, B.P. 1615,F--25010 Besan\c{c}on 
Cedex, France\\
3- D\'epartement de Physique, Observatoire du Mont M\'egantic, 
Universit\'e de Montr\'eal, Montr\'eal, Qu\'ebec, H3C 3J7 Canada}
\date{Submitted 2000 May 26}

\pagerange{\pageref{firstpage}--\pageref{lastpage}}
\pubyear{2000}

\begin{document}

\maketitle

\label{firstpage}

\begin{abstract}
We present new K--band spectroscopy for a sample of 48 starburst galaxies,
obtained using UKIRT in Hawaii. This constitutes a fair sample of the most common
types of starburst galaxies found in the nearby Universe, containing
galaxies with different morphologies, masses, metallicities and far infrared
luminosity {L$_{IR} < 10^{10}$~L$_\odot$}. The variety of near infrared spectral
features shown by these galaxies implies different bursts characteristics, which
suggests that we survey galaxies with different star formation histories or at different
stages of their burst evolution.

Using synthetic starburst models, we conclude that the
best ensemble of parameters which describe starburst galaxies in 
the nearby universe are a constant rate of star formation, a 
Salpeter IMF with an upper mass cutoff M$_{up} = 30$\ M$_\odot$ 
and bursts ages between 10 Myr and 1 Gyr. The model is fully consistent with
the differences observed in the optical and FIR between the different types of starbursts.
It suggests that H{\rm II} galaxies have younger bursts and lower metallicities than SBNGs,
while LIRGs have younger bursts but higher metallicities.

Although the above solution from the synthetic starburst model is fully
consistent with our data, it may not constitute a
strong constraint on the duration of the bursts and the IMF.
A possible alternative may be a sequence of short bursts (which may
follow an universal IMF) over a relatively long period of time.
In favour of the multiple burst hypothesis, we distinguish in our spectra some variations
of NIR features with the aperture which can be interpreted as
evidence that the burst regions are not homogeneous in space and time.
We also found that the burst stellar populations are dominated by
early--type B stars, a characteristic which seems difficult to explain with only one evolved burst.

Our observations suggest that the starburst phenomenon
must be a sustained or self--sustained phenomenon:
either star formation is continuous in time or multiple bursts happen in sequence
over a relatively long period of time. The generality of our observations
implies that this is a characteristic of starburst galaxies in the nearby Universe.

\end{abstract}


\section{Introduction}

Although a wealth of information has been accumulated on starburst galaxies
over the last 30 years, there are still fundamental questions which 
are left unanswered. In particular, we do not know if starburst galaxies form
stars according to a universal Salpeter's initial mass function (IMF), as it seems to
be the case in normal galaxies (Elmegreen 1999), or if their
IMF slope is flat or has a high lower mass cutoff, which would 
allow starburst galaxies to form preferentially massive 
stars (Rieke et al. 1993). Neither do we know if the typical 
duration of a burst is short, comparable to dynamical time--scales 
(Mas-Hesse \& Kunth 1999), or if bursts are spread over a longer period
of time (Meurer 2000).  

In the past, various observations were carried out in order to find an answer
to these two questions. Some of these studies have yield quite
surprising results. By comparing different luminosity ratios
in the optical and far infrared, Coziol (1996) found that most starburst 
galaxies may have kept forming stars at near constant rates over 
the last few billion years. The same conclusion was reached by 
Goldader et al. (1997b) for a sample of luminous infrared galaxies
(L$_{IR} > 10^{10}$ L$_\odot$). Using synthetic starburst 
models, these authors found continuous star formation and 
burst ages varying between $10^7$ and $10^9$ yr. 
They also found that the best fitted IMF is Salpeter like, but
have an upper mass cutoff M$_{up} = 30$ M$_\odot$.
This kind of IMF in starburst galaxies was previously reported 
by Doyon, Puxley \& Joseph (1992), which also suggest that the
upper mass cutoff may vary from galaxy to galaxy.

Using different arguments based on various luminosity ratios,
Deutsch \& Willner 1986, Coziol \& Demers 1995 and Coziol 1996
have found that B type stars may be predominant
in the ionised regions of massive starburst galaxies. 
Taken at face value, this last phenomenon seems consistent
with the ages of bursts deduced by Goldader et al. and 
the special form of the IMF found by these authors and
Doyon, Puxley \& Joseph. However, as pointed out by Coziol 1996, 
assuming starburst is a self--sustained phenomenon, this observation could also be interpreted in terms of a sequence of short bursts spread over a few Gyr period.

Evidence of extended or multiple bursts was recently found in many well
known active star forming galaxies, like NGC 6764 (Schinnerer, Eckart
\& Boller 2000), Arp 220 (Anantharamaiah et al. 2000), NGC 1614
(Alonso--Herrero et al. 2000) and most recently M82 (de Grijs,
O'Connell \& Gallagher 2000).
In another recent paper, Meurer (2000), using HST, distinguished
in nearby starburst galaxies different star clusters embedded 
in a diffuse glow of recently formed stars. Examining their colors,
he concluded that these clusters are consistent with
instantaneous bursts with very young ages (from 0 up to 100 Myr), 
while the diffuse light seems to be produced by stars formed continuously
over a period ten times longer than the crossing times.
A recent study of clusters in the irregular starburst galaxy NGC 1569
reveals similar clusters with ages between 3 Myrs and 1 Gyr (Hunter et al. 2000).
Old (3 Gyr) merger remnants were recently reported in the radio galaxy 
NGC 1316 (Goudfrooij et al. 2000). Multiple bursts and low upper mass cutoff for the IMF
may also be  necessary to explain the number of different Wolf--Rayet
sub--types in some WR galaxies (Schaerer et al. 2000).
Even the small mass and less evolved H{\rm II} galaxies seem
to be formed of age--composite stellar systems (Raimann et al. 2000).

In the present article, we explore these issues further
by studying the spectral characteristics in the near infrared (NIR) of a new sample
of 48 starburst galaxies with different physical characteristics. 
As it is well known, the starburst family is composed of a great variety
of galaxy types, spanning several orders in size, mass, luminosity and gas 
metallicity (Salzer, MacAlpine, \& Boroson 1989).  
The burst characteristics of these galaxies may not necessarily
be the same. By selecting galaxies with different physical 
characteristics, we therefore expect to draw a more
general picture of the starburst phenomenon. 

Our observations complete the study of luminous infrared galaxies (LIRG) made by
Goldader et al. (1995, 1997a,b). By adding these two
samples together, one can compare starburst galaxies which span
a factor 100 in B and almost 1000 in FIR luminosity.

All the galaxies in our sample have modest far infrared (FIR) luminosity
(L$_{IR} < 10^{10}$~L$_\odot$). According to the IRAS luminosity
function (Rieke \& Lebofsky 1986; Vader \& Simon 1987;
Saunders et al. 1990) such galaxies may form the bulk
of the star forming galaxies in the nearby Universe. 
Our study, therefore, should reveal the ``normal'' behaviour
in the NIR of starburst galaxies in the nearby Universe.

The K--band window in the NIR is ideal for our study.
The extinction by dust in this part of the spectrum
is $\sim 1/10$ that in the optical, which allows to compare
star formation in galaxies with very different dust content.
The K--band window also offers several diagnostic spectral
lines for studying young stellar populations: the narrow emission line
Br$\gamma$ provides some constraints on the relative number of O and B stars,
while the CO band absorption, longward of 2.3 $\mu$m, can be used to estimate
the ratio of Red Giants (RGs) over Red Super--Giants (RSGs) and better 
constrain the duration and age of the bursts (Doyon, Joseph \& Wright 1994).  
All these parameters can now be easily deduce from available
synthetic starburst models (Leitherer et al. 1999).   

\section{Presentation of the sample}

\subsection{Selection criteria and biases}

The 48 galaxies in our sample were selected using optical criteria:
they present a UV--excess in spectral lines or in the continuum, as 
established using an objective prism or the multiple filters technique. The
galaxies were taken from various sources:  the Montreal Blue Galaxy (MBG) survey
(Coziol et al. 1993; 1994), the sample of compact Kiso galaxies observed by
Augarde et al. (1994), the University of Michigan survey 
(Salzer, MacAlpine, \& Boroson 1989) and the spectrophotometric 
catalogue of H{\rm II} galaxies (Terlevich et al. 1991).

Discussion of the biases affecting these various  
surveys can be found in Coziol et al. (1997). In principle, 
by adding galaxies from objective prism and multiple filters 
surveys one can span the complete spectrum of starburst galaxies, including
H{\rm II} galaxies and Starburst Nucleus Galaxies (SBNGs).
In practice, however, observation of H{\rm II} galaxies
in the NIR is more difficult, being
less luminous than the SBNGs, the H{\rm II} galaxies need 
too long exposure times.
This explain the bias towards SBNGs in our sample.

By choosing galaxies which were 
selected based on optical criteria, we also favour galaxies which have 
a moderate to low FIR luminosity (Coziol et al. 1997).
As we already mentioned in our introduction, this bias is opportune, as
it allows us to study the most common type of starburst
galaxies found in the nearby Universe
(Rieke \& Lebofsky 1986; Vader \& Simon 1987; Saunders et al. 1990).

\begin{table*}
 \centering
 \begin{minipage}{160mm}
  \caption{Characteristics of the sample}
  \begin{tabular}{ccccccccccc} 
  {Name}      &  {cz}           &  {D}               &  {B}               &
  {Morph.}    &  {Dim.}         &  {1'}              &  {f(H$\alpha$)}    &
  {{\rm O/H}} &  {Act.}         &  { ref.}           \\ 
  { }         &  {(km s$^{-1}$)}&  {(Mpc)}           &  { }               & 
  {}          &  {(arc--min)}   &  {(kpc)}           &  {(erg s$^{-1}$)}  &
  {}          &  {Type}         &  {}\\[10pt]
MBG 00027-1645 &   7339 & 99 & 13.8 & Scd pec        &$1.4 \times 0.5$& 29 &    ...&     ...&        ...&    ...\\
MBG 00439-1342 &    675 & 10 & 13.8 & S0 pec         &$1.3 \times 1.2$&  3 &38.23  & -3.5   &SBNG       &(2b)   \\
MBG 00461-1259 &   6407 & 86 & 13.5 & (R)SB0 pec     &$0.9 \times 0.6$& 25 &41.05  & -3.7   &H{\rm II}  &(2a)   \\
UM 306         &   5096 & 70 & ...  & compact        &$< 0.5$         & 20 &    ...& -3.8   &H{\rm II}  &(3)    \\
Mrk 1002       &   3168 & 44 & 13.3 & S0             &$1.3 \times 1.2$& 13 &    ...& -3.1   &SBNG       &(8)    \\
UM 372         &  12000 &162 & ...  & compact        &$< 0.5$         & 47 &    ...& -3.9   &H{\rm II}  &(3)    \\
Mrk 363        &   2950 & 42 & 14.3 & S0 pec         &$3.8 \times 2.5$& 12 &40.38  & -3.5   &        ...&(4)    \\
MBG 02072-1022 &   3859 & 53 & 13.6 & SA(r)0 pec     &$1.1 \times 0.9$& 15 &    ...& -3.4   & SBNG      &(2a)   \\
MBG 02141-1134 &   4009 & 55 & 13.0 & Sc pec         &$1.6 \times 1.3$& 16 &40.53  & -3.1   & SBNG      &(2a)   \\
Mrk 1055       &  10830 &146 & 15.0 & compact        &$0.2 \times 0.2$& 42 &    ...&     ...&        ...&    ...\\
Mrk  602       &    849 & 13 & 13.9 & SB(rs)bc       &$1.2 \times 0.9$&  4 &40.52  &     ...&        ...&(4)    \\
KUG 0305-009   &   5860 & 80 & 15.2 & compact        &$0.3 \times 0.2$& 23 &    ...& -3.2   & SBNG      &(1)    \\
Mrk 603        &   2452 & 34 & 13.1 & S0 pec         &$1.1 \times 0.9$& 10 &    ...& -3.6   & SBNG      &(9)    \\
KUV 03073-0035 &   7080 & 96 & 17.0 & compact        &$< 0.5$         & 28 &    ...& -3.7   &H{\rm II}  &(1)    \\
MBG 03084-1059 &   5003 & 68 & 14.5 & compact        &$0.5 \times 0.5$& 20 &40.12  & -3.0   & SBNG      &(2b)   \\
MBG 03183-1853 &   3949 & 53 & 14.1 & SB(r)b         &$2.2 \times 0.4$& 16 &    ...&     ...&        ...&    ...\\
MBG 03317-2027 &   1233 & 17 & 13.5 & S0 pec         &$1.2 \times 1.1$&  5 &    ...&     ...&        ...&    ...\\
KUG 0338+032   &   6615 & 90 & 16.5 & compact        &$0.2 \times 0.1$& 26 &    ...& -3.7   &H{\rm II}  &(1)    \\
MBG 03468-2217 &   4193 & 56 & 13.8 & (R)SB(l)a      &$1.0 \times 0.6$& 16 &40.92  & -3.0   & SBNG      &(2a)   \\
MBG 03523-2034 &   1733 & 24 & 14.4 & SA0            &$1.0 \times 0.7$&  7 &    ...& -3.2   & SBNG      &(2a)   \\
IRAS 04493-0553&   2751 & 38 & 13.5 & Sb pec         &$1.3 \times 1.1$& 11 &    ...&     ...&        ...&    ...\\
Mrk 1089       &   4068 & 56 & 13.3 & SB(s)m pec/int &$0.6 \times 0.2$& 16 &    ...& -3.5   & SBNG      &(10)   \\
Mrk 1194       &   4470 & 61 & 13.4 & SB0            &$1.7 \times 1.2$& 18 &41.09  &     ...&        ...&(4)    \\
II ZW 40       &    789 & 12 & 15.5 & Sbc            &$0.6 \times 0.2$&  4 &38.95  & -4.0   & H{\rm II} &(5)    \\
KUG 0720+335   &   4028 & 56 & ...  & Sp?/int?       &$< 1.0$         & 16 &    ...& -3.0   & SBNG      &(1)    \\
Mrk 384        &   4702 & 65 & 13.8 & SBb            &$1.4 \times 0.9$& 19 &40.84  &     ...&        ...&(4)    \\
KUG 0815+249   &   2036 & 29 & 15.3 & compact        &$0.2 \times 0.2$&  9 &    ...& -3.3   & SBNG      &(1)    \\
KUG 0815+246   &   2435 & 35 & 15.3 & compact        &$0.3 \times 0.2$& 10 &    ...& -3.7   &        ...&(1)    \\
KUG 0821+229   &   7553 &103 & 15.6 & compact        &$0.3 \times 0.2$& 30 &    ...& -3.6   & SBNG      &(1)    \\
KUG 0825+252   &   2093 & 30 & 14.9 & pec/int?       &$0.7 \times 0.6$&  9 &    ...&     ...&        ...&    ...\\
Mrk 90         &   4279 & 60 & 14.1 & compact        &$0.7 \times 0.6$& 17 &40.52  &     ...&        ...&(4)    \\
Mrk 102        &   4269 & 60 & 14.5 & compact        &$0.6 \times 0.6$& 17 &40.27  & -3.7   &        ...&(4)    \\
Mrk 401        &   1699 & 25 & 13.6 & (R)SB0/a       &$1.1 \times 1.0$&  7 &40.08  &     ...&        ...&(4)    \\
I ZW 18        &    742 & 13 & 15.6 & Compact        &$< 0.5$         &  4 &    ...& -3.5   &H{\rm II}  &(6)    \\
Mrk 404        &   1266 & 19 & 12.0 & SAB(r)bc       &$2.9 \times 1.6$&  6 &    ...&     ...&        ...&    ...\\
Mrk 710        &   1494 & 22 & 13.0 & SB(r)ab        &$2.2 \times 1.4$&  6 &    ...&     ...&        ...&    ...\\
Mrk 33         &   1461 & 22 & 13.4 & Im pec         &$1.0 \times 0.9$&  6 &    ...&     ...&        ...&    ...\\
MBG 21513-1623 &  11190 &149 & 15.3 & S0/int?        &$0.2 \times 0.2$& 43 &41.41  & -3.3   & SBNG      &(2a)   \\
Mrk 307        &   5553 & 72 & 13.6 & SBc pec        &$1.1 \times 0.9$& 21 &40.73  & -3.1   &        ...&(4)    \\
MBG 22537-1650 &   3271 & 43 & 14.5 & Sab            &$0.8 \times 0.3$& 13 &40.20  & -3.4   & SBNG      &(2a)   \\
KUG 2254+124   &   7629 &100 & ...  & compact        &$< 0.5$         & 29 &    ...&     ...& SBNG      &(1)    \\
KUG 2300+163   &   2081 & 26 & 14.2 & E3 pec         &$0.9 \times 0.6$&  8 &    ...& -3.7   & H{\rm II} &(1)    \\
Mrk 326        &   3554 & 46 & 13.9 & SAB(r)bc       &$1.6 \times 1.0$& 13 &40.52  &     ...& SBNG      &(4)    \\
MBG 23318-1156 &   6180 & 82 & 14.5 & pec            &$0.8 \times 0.6$& 24 &    ...&     ...&        ...&    ...\\
Mrk 538        &   2798 & 36 & 13.0 & SB(s)b pec/int?&$1.9 \times 1.4$& 11 &    ...& -3.6   & SBNG      &(4)    \\
MBG 23372-1205 &   6429 & 85 & ...  & pec            &$1.0 \times 0.9$& 25 &40.52  & -3.5   & SBNG      &(2b)   \\
MBG 23388-1514 &   9090 &121 & 13.7 &             ...&$1.4 \times 1.3$& 35 &39.40  & -3.8   & H{\rm II} &(2b)   \\
Mrk 332        &   2406 & 30 & 13.0 & SBc            &$1.4 \times 1.3$&  9 &    ...&     ...&        ...&...           
\end{tabular}\\[10pt]
{(1) Augarde et al.\ 1994; (2a) Coziol et al.\ 1993;
(2b) Coziol et al.\ 1994; (3) Salzer, MacAlpine, \& Boroson 1989; (4) Balzano 1983; 
(5) Masegosa, Moles \& Campos-Aguilar 1994; (6) Izotov, Thuan \& Lipovetsky 1998; 
(7) Kim et al.\ 1995; (8) Veilleux \& Osterbrock 1987; (9) Liu \& Kennicutt 1995; 
(10) Mazzarella \& Boroson 1993}
\end{minipage}
\end{table*}

\subsection{Physical characteristics}

The physical characteristics of the galaxies in our sample are listed in Table~1.
The redshifts, B magnitudes, morphologies and projected dimensions in the optical
were all taken from NED\footnote{The NASA/IPAC Extragalactic Database.}.  The
distances, in column~3, were estimated assuming H$_0 = 75$ km s$^{-1}$
Mpc$^{-1}$, after correcting for the motion of the sun.  These distances were
used to determine the linear scale in kpc subtended in each galaxy by an angle
of 1 arc--min on the sky (column 7).  In many cases the morphology was not
given.  An examination of their images, as available in
NED, allows us to determine that they look compact in appearance.  These cases
are identified in column~5. Only one galaxy, MBG 23388-1514, was not
morphologically classified. Although we did not judged it compact, we could not
establish its morphology.

Various spectroscopic data were collected from the literature to complement 
our analysis. Optical spectra of these galaxies were obtained
using comparable apertures (2--3 arc--second) and median--low
spectral resolutions, allowing fair comparisons with our data.  When the
flux is available, the observed H$\alpha$ luminosity (column 8) was estimated.
Metallicities (column 9) were determined using the ratio R$_3 =
1.35\times$([O\,{\sc iii}]$\lambda5007/{\rm H}\beta)$ and the empirical relation
proposed by Vacca \& Conti (1992). For the metallicities of {\rm I}~Zw~18
and {\rm II}~Zw~40, we adopted the values estimated by Izotov, Thuan \& Lipovetsky (1998)
and Masegosa, Moles \& Campos-Aguilar (1994).  The activity types
(column 10) were established by comparing the ratios
[N{\rm II}]$\lambda6584$/H$\alpha$ with [O{\rm III}]$\lambda5007$/H$\beta$
(Baldwin, Phillips \& Terlevich 1981; Veilleux \& Osterbrock 1987).  The numbers
in the last column identify the references for the various spectroscopic data.

\begin{table}
 \centering
 \begin{minipage}{200mm}
  \caption{FIR characteristics for some galaxies in our sample}
  \begin{tabular}{ccccc}
   {Name} &   {L$_{\rm IR}$/L$_\odot$} & 
   {$\alpha$(60,25)} &   {$\alpha$(100,60)} &   {Act. type}\\[10pt]
MBG 00027-1645 &   8.18&  -2.11&  -1.77& SBNG    \\
MBG 00439-1342 &   8.61&  -2.74&  -1.47& SBNG    \\
MBG 00461-1259 &  10.22&...    &     -0.73&...      \\  
Mrk 1002       &  10.16&  -2.05&  -0.61& SBNG    \\
Mrk 363        &   9.83&...    &     -0.83&...      \\  
MBG 02072-1022 &  10.70&  -1.87&  -1.59& SBNG    \\
MBG 02141-1134 &  10.51&  -2.37&  -1.33& SBNG    \\
Mrk 1055       &  10.30&...    &     -2.32&...      \\  
Mrk  602       &   9.00&  -2.03&  -0.82& SBNG    \\
KUG 0305-009   &   9.92&...    &     -2.22&...      \\  
Mrk 603        &  10.35&  -2.04&  -0.28& SBNG    \\
MBG 03084-1059 &   9.83&...    &     -1.85&...      \\  
MBG 03183-1853 &   9.58&...    &     -2.66&...      \\  
MBG 03468-2217 &  10.29&  -1.98&  -0.47& SBNG    \\
IRAS 04493-0553&   9.83&  -2.53&  -1.34& SBNG    \\
Mrk 1194       &  10.64&  -2.57&  -1.06& SBNG    \\
Mrk 384        &  10.45&  -2.11&  -1.19& SBNG    \\
Mrk 90         &   9.83&...    &     -1.60&...      \\  
Mrk 102        &   9.64&...    &     -1.20&...      \\  
Mrk 401        &   9.42&  -1.74&  -0.87& SBNG    \\
Mrk 404        &   9.91&  -2.48&  -1.35& SBNG    \\
Mrk 710        &   9.33&  -2.16&  -0.80& SBNG    \\
MBG 21513-1623 &  10.72&  -2.13&  -1.43& SBNG    \\
Mrk 307        &  10.27&  -2.56&  -1.70& SBNG    \\
KUG 2300+163   &   9.11&...    &     -0.34&...      \\  
Mrk 326        &  10.13&  -1.95&  -0.79& SBNG    \\
MBG 23318-1156 &   9.23&...    &     -1.45&...      \\  
Mrk 538        &  10.31&  -1.47&  -0.21& SBNG    \\
MBG 23372-1205 &   9.67&...    &     -0.96&...      \\  
MBG 23388-1514 &   9.14&...    &     -1.05&...      \\  
Mrk 332        &   9.91&  -2.35&  -1.31& SBNG    \\
\end{tabular}
\end{minipage}
\end{table}

We give in Table~2 some information in the FIR for 31 galaxies (64\%) in our
sample. These data come from the IRAS Faint Source Catalogue, as found in NED.
The FIR luminosity in column 2 was determined using the relation (Londsdale et al.\
1985):  $\log({\rm L}_{\rm IR}) = \log({\rm F}_{\rm IR}) + 2 \log[z(z+1)] +
57.28$, where $z$ is the redshift, ${\rm F}_{\rm IR} = 1.26 \times 10^{-11}
(2.58 f_{60} + f_{100})$ erg cm$^{-2}$ s$^{-1}$ and $f_{60}$ and $f_{100}$ are
the fluxes in Jansky at 60 and 100 $\mu$m respectively.  In column 3 and 4, we
also give the FIR spectral indices (Sekiguchi 1987): $\alpha(\lambda_1,\lambda_2) =
\log(f_{\lambda_1}/f_{\lambda_2})/\log(\lambda_2/\lambda_1)$.  To facilitate our
analysis, only galaxies which have IRAS fluxes with high or intermediate
qualities had their FIR characteristics estimated.  The last column gives the
classification of activity type as deduced from FIR criteria (see section 2.2).

\subsection{Nature of the starburst galaxies in our sample and
physical variety of starburst galaxy hosts}

\begin{figure}
\hbox{
  \centerline{\psfig{figure=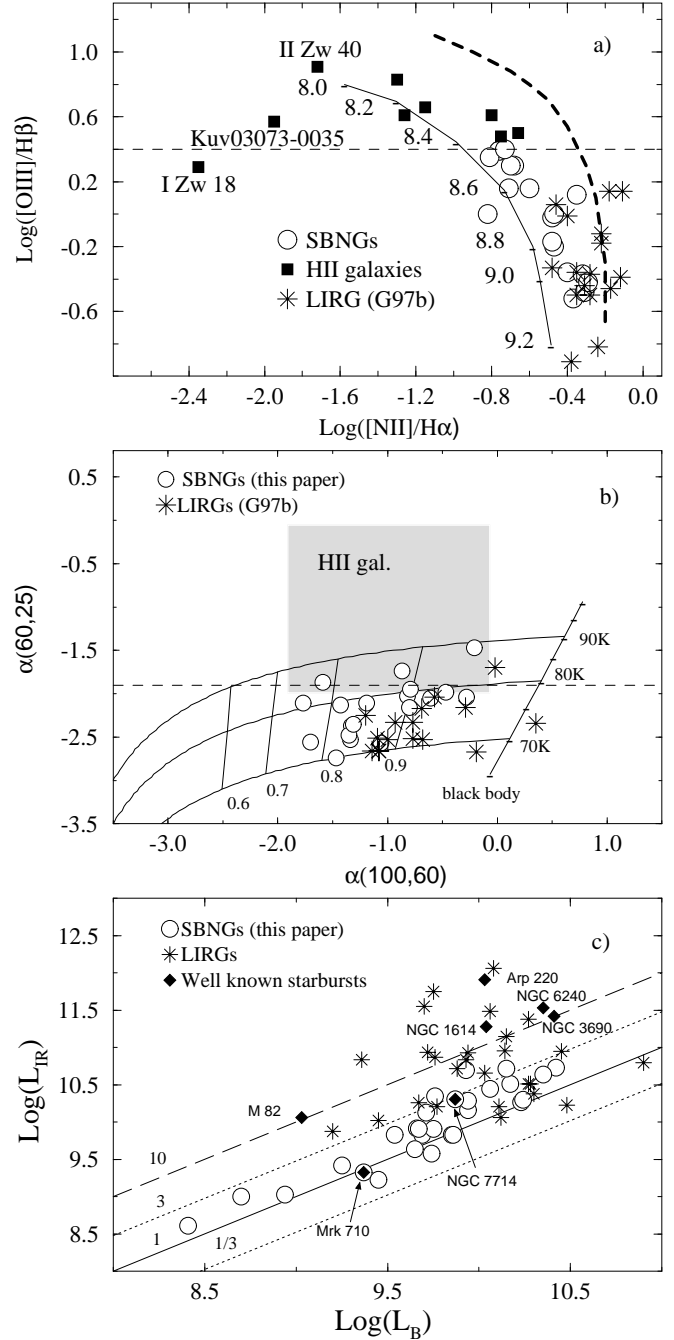,height=570pt}}}
  \caption{a) Spectroscopic diagnostic diagram in the optical. 
The bold dash curve separates starbursts from AGNs. The horizontal dash line separates
high from low excitation emission-line galaxies.
The thin continuous curve is the H{\rm II} regions sequence in metallicity 
(solar metallicity $= 8.84$ on this scale). b)
IRAS diagnostic diagram with two blackbodies model for dust.
The numbers indicate the fractional contribution from hot dust.
The horizontal dash line 
separates SBNGs from AGNs. The gray area is the locus
occupied by H{\rm II} galaxies. c) FIR vs B luminosities. The 4 diagonals correspond to 
different ratios L$_{\rm IR}$/L$_{\rm B}$}
\end{figure}

We show in Figure~1a the spectroscopic diagnostic diagram used to classify 
28 galaxies in our sample with spectroscopic data available in the optical.  
In this diagram the horizontal dash line distinguishes 
between H{\rm II} galaxies and SBNGs (Coziol 1996).
We count 19 SBNGs for 9 H{\rm II} galaxies. This result confirms our observational 
bias towards SBNGs. The thin continuous curve in Figure~1a follows the
locus traced by normal disc H{\rm II} regions (Coziol et al.\ 1994). On this curve, the gas
metallicity, given as $12 + \log({\rm O/H})$, increases as the level of excitation  
(related to [O{\rm III}]/H$\beta$) decreases. The galaxies in our sample cover the complete
sequence in starburst's metallicities.
Considering only the SBNGs, we found a mean metallicity $12 + \log({\rm O/H}) = 8.7 \pm 0.2$,
which is typical of this class of galaxies (Coziol et al.\ 1999).

Half of the galaxies that could not be classified in the optical 
have FIR characteristics which can be used to determine their starburst type:
SBNGs or H{\rm II} galaxies. In Figure~1b we show a diagnostic diagram based
on IRAS spectral indices $\alpha$(60,25) and $\alpha$(100,60).
Starburst galaxies are described in this diagram by a model with 
two dust components (Sekiguchi 1987): a cold component with a dust temperature 
of 27 K and a hot component with dust temperatures varying between 70 and 90 K. 
According to this model, starburst galaxies differ from normal spiral galaxies by their
hot dust contribution, which is usually higher than 60\%. Although very few 
H{\rm II} galaxies are detected in the FIR, they can usually be distinguished
from SBNGs from their hotter dust temperatures. 
This is indicated by the gray area in Figure~1b.
According to this last criterion, most of the galaxies with FIR data 
in our sample are SBNGs.

\begin{table*}
 \centering
 \begin{minipage}{500mm}
  \caption{Fluxes and equivalent widths for Br$\gamma$ and H$_2$ lines}
  \begin{tabular}{cccccccccccccccccc} 
    { }    &    { } & \multicolumn{4}{c}{Aperture 1} &\multicolumn{4}{c}{Aperture 2}\\ 
    { }    &    { } & \multicolumn{4}{c}{--------------------------------------------------------} &\multicolumn{4}{c}{--------------------------------------------------------}\\ 
    {Name} &    {z} & {f(Br$\gamma$)}    & {EW(Br$\gamma$)} & {f(H$_2$)}            & {EW(H$_2$)}
                    & {f(Br$\gamma$)}    & {EW(Br$\gamma$)} & {f(H$_2$)}            & {EW(H$_2$)}  \\
    { }    &    { } & {$\times 10^{-15}$}&    {}            &    {$\times 10^{-15}$}&    {}
                    & {$\times 10^{-15}$}&    {}            &    {$\times 10^{-15}$}&    {}        \\
    { }    &    { } & {(erg s$^{-1}$)}   &    {(\AA)}       &    {(erg s$^{-1}$)}   &    {(\AA)}
                    & {(erg s$^{-1}$)}   &    {(\AA)}       &    {(erg s$^{-1}$)}   &    {(\AA)}   \\[10pt]
MBG00027-1645 & 0.0269 &    2.1\ $\pm$\  0.3&   4.1\ $\pm$\  0.7&  1.7\ $\pm$\  0.3  &  3.2\ $\pm$\  0.6 
                       &    0.7\ $\pm$\  0.2&   2.6\ $\pm$\  0.8&    $<$ 0.8         &    $<$ 2.9        \\
MBG00439-1342 & 0.0019 &  $<$ 2.7           &     $<$ 3.7       &    $<$ 1.6         &    $<$ 2.1        
                       &  $<$ 0.9           &     $<$ 2.7       &    $<$ 0.7         &    $<$ 2.1        \\
MBG00461-1259 & 0.0232 &    1.8\ $\pm$\  0.4&   11.0\ $\pm$\ 2.7&    $<$ 0.8         &    $<$ 4.5        
                       &    0.7\ $\pm$\  0.2&   8.3\ $\pm$\  1.7&    $<$ 0.3         &    $<$ 3.2        \\
UM306         & 0.0175 &  $<$ 0.2           &   $<$ 10.9        &    $<$ 0.2         &    $<$ 7.2        
                       &  $<$ 0.1           &   $<$ 11.2        &    $<$ 0.1         &    $<$ 11.6       \\
Mrk1002       & 0.0121 &    8.8\ $\pm$\  0.6&   8.4\ $\pm$\  0.6&  2.7\ $\pm$\  0.8  &  2.5\ $\pm$\  0.7 
                       &    5.5\ $\pm$\  0.4&   8.6\ $\pm$\  0.6&    $<$ 1.3         &    $<$ 1.9        \\
UM372         & 0.0395 &  $<$ 0.9           &   $<$ 11.7        &    $<$ 0.7         &    $<$  8.4       
                       &  $<$ 0.4           &   $<$ 5.4         &    $<$ 0.5         &    $<$  6.7       \\
Mrk363        & 0.0087 &  $<$ 1.0           &   $<$ 3.0         &    $<$ 1.6         &    $<$ 4.4        
                       &  $<$ 0.4           &   $<$ 2.7         &    $<$ 0.6         &    $<$ 3.8        \\
MBG02072-1022 & 0.0138 &   37.4\ $\pm$\  1.9&  10.5\ $\pm$\  0.5&    $<$ 11.5        &    $<$  3.1       
                       &   25.2\ $\pm$\  0.8&  13.9\ $\pm$\  0.4&    $<$ 5.4         &    $<$  2.8       \\
MBG02141-1134 & 0.0149 &    4.2\ $\pm$\  0.5&   7.7\ $\pm$\  0.9&  2.0\ $\pm$\  0.4  &  3.5\ $\pm$\  0.6 
                       &    1.7\ $\pm$\  0.4&   7.6\ $\pm$\  1.7&  0.8\ $\pm$\  0.2  &  3.5\ $\pm$\  0.8 \\
Mrk1055       & 0.0388 &  $<$ 1.1           &    $<$ 4.2        &    $<$  1.1        &    $<$  3.7       
                       &  $<$ 0.8           &    $<$ 4.1        &    $<$  0.6        &    $<$  2.8       \\
MRK602        & 0.0018 &  $<$ 6.1           &    $<$ 7.3        &    $<$  3.0        &    $<$  3.5       
                       &  $<$ 5.5           &    $<$ 9.1        &    $<$  2.2        &    $<$  3.5       \\
KUG0305-009   & 0.0214 &    1.1\ $\pm$\  0.3&   4.2\ $\pm$\  1.3&    $<$  0.8        &    $<$  2.8       
                       &    0.6\ $\pm$\  0.2&   3.9\ $\pm$\  1.2&    $<$  0.6        &    $<$  3.8       \\
Mrk603        & 0.0095 &   23.5\ $\pm$\  0.8&  22.9\ $\pm$\  0.8&  4.1\ $\pm$\  0.8  &  3.7\ $\pm$\  0.8 
                       &   16.9\ $\pm$\  0.4&  23.4\ $\pm$\  0.6&  1.4\ $\pm$\  0.3  &  1.8\ $\pm$\  0.4 \\
KUV03073-0035 & 0.0236 &    1.0\ $\pm$\  0.3&   5.0\ $\pm$\  1.4&    $<$  1.0        &    $<$ 4.5        
                       &    0.6\ $\pm$\  0.1&   5.6\ $\pm$\  0.9&    $<$  0.6        &    $<$ 5.1        \\
MBG03084-1059 & 0.0180 &  $<$ 1.2           &    $<$ 4.3        &    $<$  0.9        &    $<$  3.3       
                       &    0.8\ $\pm$\  0.2&   5.4\ $\pm$\  1.1&    $<$  0.5        &    $<$  3.3       \\
MBG03183-1853 & 0.0132 &  $<$ 0.7           &    $<$ 3.0        &    $<$  1.1        &    $<$  4.4       
                       &  $<$ 0.4           &    $<$ 3.5        &    $<$  0.4        &    $<$  3.8       \\
MBG03317-2027 & 0.0047 &  $<$ 1.6           &    $<$ 2.1        &    $<$  2.8        &    $<$  3.5       
                       &  $<$ 1.4           &    $<$ 2.8        &    $<$  1.3        &    $<$  2.4       \\
KUG0338+032   & 0.0228 &  $<$ 0.7           &    $<$ 8.3        &    $<$  0.8        &    $<$  8.2       
                       &    0.4\ $\pm$\  0.1&   5.8\ $\pm$\  1.8&  $<$0.3        &   $<$  4.4        \\
MBG03468-2217 & 0.0157 &    8.6\ $\pm$\  0.7&   8.4\ $\pm$\  0.7&  3.3\ $\pm$\  0.5&  3.1\ $\pm$\  0.4
                       &    6.3\ $\pm$\  0.5&   8.2\ $\pm$\  0.6&  1.6\ $\pm$\  0.4&  2.0\ $\pm$\  0.5\\
MBG03523-2034 & 0.0070 &    1.3\ $\pm$\  0.4&   5.4\ $\pm$\  1.7&  1.1\ $\pm$\  0.3&  4.7\ $\pm$\  1.1
                       &  $<$ 0.5           &    $<$ 3.7        &    $<$ 0.3        &    $<$ 2.0        \\
IRAS04493-0553& 0.0095 &    1.6\ $\pm$\  0.3&  7.4\ $\pm$\  1.4 &  0.9\ $\pm$\  0.3&  3.9\ $\pm$\  1.2
                       &    0.5\ $\pm$\  0.2&  6.9\ $\pm$\  2.3 &  $<$ 0.4        &   $<$ 5.5        \\
Mrk1089       & 0.0148 &   10.7\ $\pm$\  0.3& 84.9\ $\pm$\  3.9 &  1.8\ $\pm$\  0.2&  13.5\ $\pm$\  1.9
                       &    3.7\ $\pm$\  0.2& 79.0\ $\pm$\  4.7 &  0.9\ $\pm$\  0.2&  18.8\ $\pm$\  4.5\\
Mrk1194       & 0.0167 &    5.6\ $\pm$\  0.5&  4.1\ $\pm$\  0.4 &  3.7\ $\pm$\  0.6&  2.6\ $\pm$\  0.4
                       &    1.1\ $\pm$\  0.3&  1.8\ $\pm$\  0.5 &  1.5\ $\pm$\  0.3&  2.3\ $\pm$\  0.4\\
IIZW40        & 0.0034 &   45.1\ $\pm$\  0.4& 264.4\ $\pm$\ 9.3 &  2.3\ $\pm$\  0.5&  13.0\ $\pm$\  2.7
                       &   34.3\ $\pm$\  0.4& 322.0\ $\pm$\ 15.0&  1.2\ $\pm$\  0.2&  11.6\ $\pm$\  1.4\\
KUG0720+335   & 0.0134 &   16.9\ $\pm$\  0.6&  18.4\ $\pm$\  0.7&  3.5\ $\pm$\  0.7&  3.6\ $\pm$\  0.8
                       &   13.2\ $\pm$\  0.4&  18.9\ $\pm$\  0.6&  2.1\ $\pm$\  0.4&  2.8\ $\pm$\  0.5\\
Mrk384        & 0.0155 &    3.0\ $\pm$\  1.0&  3.7\ $\pm$\  1.2 &   $<$  3.1        &    $<$  3.7        
                       &    2.0\ $\pm$\  0.4&  5.0\ $\pm$\  0.9 &   $<$  1.5        &    $<$  3.5        \\
KUG0815+249   & 0.0075 &    1.5\ $\pm$\  0.4&  7.7\ $\pm$\  1.9 &   $<$  1.0        &    $<$  4.8        
                       &    0.6\ $\pm$\  0.1&  5.8\ $\pm$\  1.1 &   $<$  0.7        &    $<$  6.2        \\
KUG0815+246   & 0.0091 &  $<$  0.9          &    $<$ 7.4        &   $<$ 1.4        &    $<$ 10.4        
                       &  $<$  0.6          &    $<$ 8.5        &   $<$ 0.7        &    $<$  8.7        \\
KUG0821+229   & 0.0253 &    0.5\ $\pm$\  0.2&  4.2\ $\pm$\  1.4 &   $<$  0.9        &    $<$  7.1        
                       &    0.5\ $\pm$\  0.2&  8.1\ $\pm$\  2.5 &   $<$  0.8        &   $<$  11.1        \\
KUG0825+252   & 0.0071 &  $<$  0.4          &    $<$  3.9&   $<$  0.7        &   $<$   6.5        
                       &  $<$  0.3          &    $<$  5.5&   $<$  0.3        &    $<$  4.7        \\
Mrk90         & 0.0112 &  $<$  2.0          &    $<$  4.8&   $<$  1.0        &    $<$  2.4        
                       &  $<$  0.8          &    $<$  4.2&   $<$  1.0        &    $<$  4.7        \\
Mrk102        & 0.0105 &  $<$  2.0          &    $<$  3.5&   $<$  2.5        &    $<$  4.1        
                       &    1.2\ $\pm$\  0.4&  4.4\ $\pm$\  1.5&   $<$  1.8        &    $<$  6.2        \\
Mrk401        & 0.0061 &   6.8\ $\pm$\  0.7&  5.2\ $\pm$\  0.5&   $<$  2.4        &    $<$  1.7        
                       &  6.4\ $\pm$\  0.5&  7.3\ $\pm$\  0.6&   $<$  1.5        &    $<$  1.6        \\
IZW18         & 0.0020 &    $<$  2.0    &   $<$  180.4&   $<$  1.3        &   $<$  85.3        
                       &    $<$  0.8    &   $<$  76.5&   $<$  0.7        &   $<$  70.4        \\
Mrk404        & 0.0038 &   11.5\ $\pm$\  1.3&  5.3\ $\pm$\  0.6&   $<$  4.4        &    $<$  1.9        
                       &    8.9\ $\pm$\  0.7&  6.1\ $\pm$\  0.5&  3.0\ $\pm$\  1.0&  2.0\ $\pm$\  0.6\\
Mrk710        & 0.0055 &   18.3\ $\pm$\  0.6&  36.3\ $\pm$\  1.2&  3.1\ $\pm$\  0.6&  5.7\ $\pm$\  1.1
                       &  15.1\ $\pm$\  0.4&  57.5\ $\pm$\  2.2&  2.3\ $\pm$\  0.3&  8.3\ $\pm$\  1.2\\
Mrk33         & 0.0040 &   29.8\ $\pm$\  0.9&  34.0\ $\pm$\  1.1&  3.2\ $\pm$\  0.8&  3.5\ $\pm$\  0.9
                       &  18.6\ $\pm$\  0.5&  33.8\ $\pm$\  1.0&  2.0\ $\pm$\  0.4&  3.5\ $\pm$\  0.7\\
MBG21513-1623 & 0.0381 &   0.5\ $\pm$\  0.1&  6.5\ $\pm$\  1.5&   $<$  0.6        &    $<$  7.0        
                       &    $<$  0.4    &   $<$  11.3&   $<$  0.3        &    $<$  7.1        \\
Mrk307        & 0.0193 &    $<$  2.1    &    $<$  8.9&   $<$  1.4        &    $<$  5.4        
                       &    1.2\ $\pm$\  0.3&  10.4\ $\pm$\  2.3&   $<$  0.8        &    $<$  6.3        \\
MBG22537-1650 & 0.0138 &   1.0\ $\pm$\  0.3&  3.5\ $\pm$\  1.0&   $<$  1.2        &    $<$  4.0        
                       &    0.6\ $\pm$\  0.2&  4.0\ $\pm$\  1.0&   $<$  0.6        &    $<$  3.5        \\
KUG2254+124   & 0.0256 &   0.8\ $\pm$\  0.2&  10.6\ $\pm$\  2.7&   $<$  0.8        &    $<$  9.9        
                       &    0.4\ $\pm$\  0.1&  8.8\ $\pm$\  2.8&   $<$  0.5        &    $<$ 10.8        \\
KUG2300+163   & 0.0082 &   1.3\ $\pm$\  0.2&  21.4\ $\pm$\  2.5&   $<$  0.5        &    $<$  7.0        
                       &    0.6\ $\pm$\  0.1&  22.0\ $\pm$\  2.8&   $<$  0.2        &    $<$  5.9        \\
Mrk326        & 0.0140 &   5.2\ $\pm$\  0.6&  5.3\ $\pm$\  0.6&   $<$  2.4        &    $<$  2.3        
                       &  6.1\ $\pm$\  0.4&  8.4\ $\pm$\  0.6&  2.4\ $\pm$\  0.6&  3.1\ $\pm$\  0.8\\
MBG23318-1156 & 0.0206 &   0.9\ $\pm$\  0.2&  12.9\ $\pm$\  3.0&   $<$  1.0        &   $<$  12.4        
                       &    0.5\ $\pm$\  0.2&  15.4\ $\pm$\  4.9&   $<$  0.4        &   $<$  10.1        \\
Mrk538        & 0.0110 &   14.6\ $\pm$\  0.9&  17.1\ $\pm$\  1.1&  6.7\ $\pm$\  1.0&  7.5\ $\pm$\  1.1
                       & 10.5\ $\pm$\  0.4&  18.9\ $\pm$\  0.7&  3.2\ $\pm$\  0.3&  5.4\ $\pm$\  0.6\\
MBG23372-1205 & 0.0239 &   0.8\ $\pm$\  0.2&  9.0\ $\pm$\  2.7&   $<$  0.7        &    $<$  7.6        
                       &  $<$  0.9    &   $<$  21.3&   $<$  0.8        &   $<$  16.8        \\
MBG23388-1514 & 0.0303 &    $<$  1.6    &   $<$  23.3&   $<$  1.6        &   $<$  23.0        
                       &  $<$  0.5    &   $<$  13.7&   $<$  0.6        &   $<$  14.8        \\
Mrk332        & 0.0110 &   2.5\ $\pm$\  0.8&  2.6\ $\pm$\  0.8&  2.6\ $\pm$\  0.7&  2.6\ $\pm$\  0.7
                       &  2.2\ $\pm$\  0.6&  3.7\ $\pm$\  0.9&  1.9\ $\pm$\  0.5&  2.9\ $\pm$\  0.8
\end{tabular}
\end{minipage}
\end{table*}

\begin{table*}
 \centering
 \begin{minipage}{160mm}
  \caption{Continuum slope, CO index and K magnitudes}
  \begin{tabular}{ccccccccccc}
    { }    &    { } & \multicolumn{4}{c}{Aperture 1} &\multicolumn{4}{c}{Aperture 2}\\
    { }    &    { } & \multicolumn{4}{c}{--------------------------------------------------------} &\multicolumn{4}{c}{--------------------------------------------------------}\\ 
    {Name} &    {z} &    {$\beta$} &    {CO$_{\rm spec}$} &    {m$_{\rm K}$} &    {S/N}
                    &    {$\beta$} &    {CO$_{\rm spec}$} &    {m$_{\rm K}$} &    {S/N} \\  
    { }    &    {}  &    {}        &    {(mag.)}          &    {}            &    {}
                    &    {}        &    {(mag.)}          &    {}            &    {}\\[10pt]
MBG00027-1645 & 0.0269 &-2.85\ $\pm$\  0.14 & 0.21\ $\pm$\  0.01 &12.32   &116 
                       &-3.01\ $\pm$\  0.14 & 0.25\ $\pm$\  0.01 &13.04   & 92 \\
MBG00439-1342 & 0.0019 &-3.15\ $\pm$\  0.13 & 0.23\ $\pm$\  0.01 &12.14   &115 
&-2.90\ $\pm$\  0.10 & 0.20\ $\pm$\  0.01 &12.99   & 92 \\
MBG00461-1259 & 0.0232 &-3.33\ $\pm$\  0.23 & 0.17\ $\pm$\  0.02 &13.55   & 49 
&-3.27\ $\pm$\  0.23 & 0.17\ $\pm$\  0.02 &14.23   & 48 \\
UM306         & 0.0175 &-4.37\ $\pm$\  0.35 &-0.02\ $\pm$\  0.04 &15.74   & 26 
&-3.66\ $\pm$\  0.44 & 0.03\ $\pm$\  0.05 &16.49   & 23 \\
Mrk1002       & 0.0121 &-2.40\ $\pm$\  0.10 & 0.19\ $\pm$\  0.01 &11.59   &101 
&-2.03\ $\pm$\  0.08 & 0.17\ $\pm$\  0.01 &12.09   &111 \\
UM372         & 0.0395 &-3.11\ $\pm$\  0.34 & 0.02\ $\pm$\  0.03 &14.22   & 23 
&-3.84\ $\pm$\  0.25 & 0.03\ $\pm$\  0.03 &14.32   & 50 \\
Mrk363        & 0.0087 &-3.08\ $\pm$\  0.10 & 0.15\ $\pm$\  0.01 &12.85   &108 
&-3.01\ $\pm$\  0.12 & 0.14\ $\pm$\  0.01 &13.72   & 92 \\
MBG02072-1022 & 0.0138 &-2.80\ $\pm$\  0.13 & 0.29\ $\pm$\  0.01 &10.34   & 53 
&-3.06\ $\pm$\  0.12 & 0.32\ $\pm$\  0.01 &11.09   &272 \\
MBG02141-1134 & 0.0149 &-3.01\ $\pm$\  0.12 & 0.20\ $\pm$\  0.01 &12.34   & 85 
&-2.56\ $\pm$\  0.13 & 0.15\ $\pm$\  0.01 &13.26   & 54 \\
Mrk1055       & 0.0388 &-3.61\ $\pm$\  0.21 & 0.16\ $\pm$\  0.02 &12.95   & 49 
&-4.01\ $\pm$\  0.21 & 0.21\ $\pm$\  0.02 &13.33   & 60 \\
MRK602        & 0.0018 &-2.97\ $\pm$\  0.17 & 0.22\ $\pm$\  0.01 &11.93   &132 
&-2.93\ $\pm$\  0.19 & 0.20\ $\pm$\  0.02 &12.27   &185 \\
KUG0305-009   & 0.0214 &-2.90\ $\pm$\  0.12 & 0.17\ $\pm$\  0.01 &13.12   & 70 
&-2.31\ $\pm$\  0.13 & 0.15\ $\pm$\  0.01 &13.70   & 73 \\
Mrk603        & 0.0095 &-2.83\ $\pm$\  0.08 & 0.16\ $\pm$\  0.01 &11.61   &107 
&-2.79\ $\pm$\  0.07 & 0.17\ $\pm$\  0.01 &11.99   &187 \\
KUV03073-0035 & 0.0236 &-2.49\ $\pm$\  0.14 & 0.23\ $\pm$\  0.02 &13.28   & 73 
&-2.50\ $\pm$\  0.14 & 0.26\ $\pm$\  0.02 &13.94   & 68 \\
MBG03084-1059 & 0.0180 &-2.74\ $\pm$\  0.14 & 0.20\ $\pm$\  0.01 &13.02   & 63 
&-2.95\ $\pm$\  0.15 & 0.25\ $\pm$\  0.01 &13.73   & 52 \\
MBG03183-1853 & 0.0132 &-3.24\ $\pm$\  0.12 & 0.16\ $\pm$\  0.01 &13.19   & 67 
&-3.35\ $\pm$\  0.16 & 0.15\ $\pm$\  0.01 &14.17   & 55 \\
MBG03317-2027 & 0.0047 &-3.03\ $\pm$\  0.10 & 0.22\ $\pm$\  0.01 &12.01   & 80 
&-2.90\ $\pm$\  0.10 & 0.21\ $\pm$\  0.01 &12.43   & 90 \\
KUG0338+032   & 0.0228 &-3.11\ $\pm$\  0.23 & 0.12\ $\pm$\  0.02 &14.24   & 30 
&-3.40\ $\pm$\  0.21 & 0.12\ $\pm$\  0.03 &14.53   & 46 \\
MBG03468-2217 & 0.0157 &-2.36\ $\pm$\  0.10 & 0.27\ $\pm$\  0.01 &11.61   & 85 
&-2.51\ $\pm$\  0.10 & 0.28\ $\pm$\  0.01 &11.93   &129 \\
MBG03523-2034 & 0.0070 &-2.60\ $\pm$\  0.18 & 0.21\ $\pm$\  0.02 &13.22   & 42 
&-2.40\ $\pm$\  0.15 & 0.23\ $\pm$\  0.02 &13.90   & 49 \\
IRAS04493-0553& 0.0095 &-3.04\ $\pm$\  0.15 & 0.18\ $\pm$\  0.02 &13.35   & 37 
&-2.49\ $\pm$\  0.23 & 0.16\ $\pm$\  0.03 &14.50   & 21 \\
Mrk1089       & 0.0148 &-2.45\ $\pm$\  0.32 & 0.19\ $\pm$\  0.04 &13.83   & 26 
&-3.68\ $\pm$\  0.45 & 0.16\ $\pm$\  0.05 &14.97   & 18 \\
Mrk1194       & 0.0167 &-2.66\ $\pm$\  0.09 & 0.22\ $\pm$\  0.01 &11.33   &240 
&-2.67\ $\pm$\  0.09 & 0.21\ $\pm$\  0.01 &12.22   &222 \\
IIZW40        & 0.0034 &-1.97\ $\pm$\  0.15 & 0.07\ $\pm$\  0.01 &13.42   & 64 
&-1.79\ $\pm$\  0.18 & 0.06\ $\pm$\  0.02 &13.91   & 73 \\
KUG0720+335   & 0.0134 &-2.66\ $\pm$\  0.08 & 0.24\ $\pm$\  0.01 &11.72   &120 
&-2.75\ $\pm$\  0.08 & 0.26\ $\pm$\  0.01 &12.02   &169 \\
Mrk384        & 0.0155 &-3.11\ $\pm$\  0.12 & 0.25\ $\pm$\  0.01 &11.91   & 52 
&-2.13\ $\pm$\  0.12 & 0.16\ $\pm$\  0.01 &12.60   & 42 \\
KUG0815+249   & 0.0075 &-3.19\ $\pm$\  0.21 & 0.16\ $\pm$\  0.02 &13.40   & 51 
&-3.90\ $\pm$\  0.20 & 0.16\ $\pm$\  0.02 &14.09   & 49 \\
KUG0815+246   & 0.0091 &-3.20\ $\pm$\  0.24 & 0.14\ $\pm$\  0.02 &13.91   & 35 
&-3.45\ $\pm$\  0.31 & 0.12\ $\pm$\  0.04 &14.55   & 35 \\
KUG0821+229   & 0.0253 &-3.28\ $\pm$\  0.30 & 0.11\ $\pm$\  0.03 &13.87   & 34 
&-4.25\ $\pm$\  0.27 & 0.14\ $\pm$\  0.02 &14.57   & 33 \\
KUG0825+252   & 0.0071 &-3.06\ $\pm$\  0.17 & 0.18\ $\pm$\  0.01 &14.19   & 36 
&-3.56\ $\pm$\  0.21 & 0.15\ $\pm$\  0.01 &14.89   & 36 \\
Mrk90         & 0.0112 &-3.24\ $\pm$\  0.18 & 0.22\ $\pm$\  0.02 &12.66   & 58 
&-2.98\ $\pm$\  0.13 & 0.16\ $\pm$\  0.01 &13.41   & 50 \\
Mrk102        & 0.0105 &-3.68\ $\pm$\  0.14 & 0.18\ $\pm$\  0.01 &12.29   & 71 
&-3.16\ $\pm$\  0.16 & 0.12\ $\pm$\  0.01 &13.06   & 55 \\
Mrk401        & 0.0061 &-2.91\ $\pm$\  0.08 & 0.21\ $\pm$\  0.01 &11.43   &286 
&-2.80\ $\pm$\  0.08 & 0.22\ $\pm$\  0.01 &11.86   &306 \\
IZW18         & 0.0020 &-3.63\ $\pm$\  3.36 & 0.43\ $\pm$\  0.34 &16.45   &   2 
&-6.69\ $\pm$\  2.88 &-0.01\ $\pm$\  0.29 &16.89   &   3 \\
Mrk404        & 0.0038 &-2.69\ $\pm$\  0.11 & 0.24\ $\pm$\  0.01 &10.85   & 91 
&-2.38\ $\pm$\  0.08 & 0.23\ $\pm$\  0.01 &11.26   & 88 \\
Mrk710        & 0.0055 &-2.06\ $\pm$\  0.22 & 0.09\ $\pm$\  0.02 &12.33   & 64 
&-1.65\ $\pm$\  0.24 & 0.10\ $\pm$\  0.03 &13.02   & 71 \\
Mrk33         & 0.0040 &-2.63\ $\pm$\  0.10 & 0.20\ $\pm$\  0.01 &11.80   & 95 
&-2.57\ $\pm$\  0.08 & 0.19\ $\pm$\  0.01 &12.31   &117 \\
MBG21513-1623 & 0.0381 &-2.60\ $\pm$\  0.26 & 0.14\ $\pm$\  0.02 &14.26   & 32 
&-3.18\ $\pm$\  0.45 & 0.12\ $\pm$\  0.05 &15.22   & 23 \\
Mrk307        & 0.0193 &-3.24\ $\pm$\  0.23 & 0.21\ $\pm$\  0.02 &13.19   & 49 
&-2.98\ $\pm$\  0.22 & 0.24\ $\pm$\  0.02 &13.96   & 38 \\
MBG22537-1650 & 0.0138 &-2.70\ $\pm$\  0.11 & 0.16\ $\pm$\  0.01 &12.92   & 78 
&-2.88\ $\pm$\  0.13 & 0.17\ $\pm$\  0.01 &13.60   & 92 \\
KUG2254+124   & 0.0256 &-3.84\ $\pm$\  0.25 & 0.10\ $\pm$\  0.03 &14.33   & 25 
&-4.65\ $\pm$\  0.31 & 0.20\ $\pm$\  0.04 &14.98   & 31 \\
KUG2300+163   & 0.0082 &-2.54\ $\pm$\  0.22 & 0.19\ $\pm$\  0.02 &14.61   & 44 
&-2.49\ $\pm$\  0.21 & 0.17\ $\pm$\  0.02 &15.58   & 52 \\
Mrk326        & 0.0140 &-2.87\ $\pm$\  0.10 & 0.24\ $\pm$\  0.01 &11.67   &105 
&-2.68\ $\pm$\  0.10 & 0.27\ $\pm$\  0.01 &12.00   &166 \\
MBG23318-1156 & 0.0206 &-3.67\ $\pm$\  0.34 & 0.05\ $\pm$\  0.03 &14.44   & 24 
&-3.59\ $\pm$\  0.34 & 0.11\ $\pm$\  0.04 &15.29   & 19 \\
Mrk538        & 0.0110 &-3.11\ $\pm$\  0.13 & 0.22\ $\pm$\  0.01 &11.83   & 82 
&-3.19\ $\pm$\  0.13 & 0.27\ $\pm$\  0.01 &12.31   & 79 \\
MBG23372-1205 & 0.0239 &-3.28\ $\pm$\  0.49 & 0.09\ $\pm$\  0.05 &14.22   & 22 
&-3.16\ $\pm$\  0.75 & 0.07\ $\pm$\  0.08 &15.04   & 18 \\
MBG23388-1514 & 0.0303 &-5.80\ $\pm$\  0.78 &-0.12\ $\pm$\  0.08 &14.59   & 13 
&-5.33\ $\pm$\  0.57 &-0.03\ $\pm$\  0.07 &15.30   & 13 \\
Mrk332        & 0.0110 &-3.51\ $\pm$\  0.15 & 0.21\ $\pm$\  0.01 &11.75   &129 
&-3.45\ $\pm$\  0.15 & 0.23\ $\pm$\  0.01 &12.22   &153 
\end{tabular}
\end{minipage}
\end{table*}

\subsection{Probability to find an AGN in our sample}

Using FIR criteria we can also ascertain what is the probability that one
of the galaxies in our sample is dominated by an AGN. 
According to Coziol et al.\ (1998), 99\% of IRAS starburst galaxies 
have $\alpha$(60,25) $<-1.9$ (the horizontal dash line 
in Figure~1b), while only 25\% of Seyfert galaxies have such FIR colours.
Almost all the galaxies in our sample have $\alpha$(60,25) $<-1.9$,
which excludes the possibility
that one of these galaxies is dominated by an AGN. 
Note that the three galaxies exceeding this limit in our sample
(Mrk 538, Mrk 401 and MBG 02072-1022)
have already been classified as SBNGs from their spectra. 
In fact, all the galaxies previously classified as starburst based
on their optical spectra occupy
this region of the diagram, which thus confirms their nature as starburst.

Only six of the galaxies detected in the FIR diagram do not have a spectral 
classification in the optical. Based on the reported statistics above, 
and the total number of galaxies which was necessary to established it
(see Coziol et al.\ 1998), we estimate to 7\% the probability to find
an AGN (which is probably not predominant) in this region of the diagram.
This reduce to none the probability that one of these six galaxies also host an AGN.
This results is not surprising considering the moderate FIR luminosity of
these galaxies and the fact that the probability to find
an AGN increases with the FIR luminosity (Veilleux, Kim \& Sanders 1999).

The above conclusion hold also for the ten galaxies in our sample without spectral 
classification or FIR detection. Height of these galaxies were classified
as starburst based on an objective prism spectrum, which rules out the
presence of a broad line or luminous AGN. The
last two are MBG galaxies which were classified as starburst based on 
narrow lines and small [N{\rm II}]/H$\alpha$ ratios in one spectrum. 
The fact that these galaxies were not detected in IRAS suggests
that the probability they host an AGN is practically null.

\subsection{Comparison with LIRGs and uLIRGs}

At this point of our analysis, it is convenient to compare 
our sample of galaxies with the 16 LIRGs
previously classified as starburst in the study of Goldader et al.\ (1997a).  
In Figure~1a, we see that the LIRGs have spectroscopical
characteristics similar to SBNGs. However, they are generally more metal rich, having
solar metallicity  on average. Their hot dust contribution (Figure~1b) 
is also higher than in SBNGs, which is consistent with 
their higher FIR luminosity.      

In Figure~1c, we compare the FIR and B luminosities of our galaxies 
with those of the LIRGs. For comparison, we also
include some ultra luminous infrared galaxies (uLIRGs) from the literature.
The ratios L$_{\rm  IR}$/L$_{\rm B}$
can be used to distinguish galaxies with unusually high star formation rates
(see Coziol 1996): normal late--type spiral galaxies with constant
star formation rates over the last 3 Gyrs have $1/3 < $L$_{\rm  IR}$/L$_{\rm B}$ $< 3$.
One can see that the ratio L$_{\rm  IR}$/L$_{\rm B}$ tends to increase
with the B luminosity. The uLIRGs and LIRGs have both higher
B luminosities and, therefore, higher L$_{\rm  IR}$/L$_{\rm B}$ ratios as compared to SBNGs.
Most of the galaxies in our sample have ratios L$_{\rm  IR}$/L$_{\rm B}$
varying between 1 and 3. This comparison suggests that uLIRGs and LIRGs
are generally more massive than SBNGs (having
higher B luminosity) and may have more active burst regions or have younger bursts
(as judged from their higher L$_{\rm  IR}$/L$_{\rm B}$ ratios).

One remarkable feature of Figure~1c is how the starburst galaxies in our sample
differ from well known uLIRGs, like Arp~220, NGC~6240, NGC~1614 or NGC~3690.
The nature of uLIRGs is still in debate. It is not sure if part of the activity
observed in these galaxies is not due to an hidden AGN. 
This is not the case of the galaxies in our sample. The position, in Figure~1c, of
the ''prototype`` starburst galaxy M82
seems, however, interesting. This galaxy looks more like an uLIRG than to one of our
galaxies. This is consistent with the extreme properties observed in
this object, which may also suggest some kind of AGN activity
(Seaquist, Frayer \& Frail 1997; Allen \& Kronberg 1998; Kaaret et al. 2000).
Two other archetype starburst galaxies in our sample are
NGC~7714 (Mrk 538) and Mrk~710. These two galaxies are more representative of
the starburst activity in our sample of galaxies.

The above analysis emphasizes the diversity of physical characteristics
of galaxies hosting a starburst. Our observations should reveal if this
variety of physical characteristics is reflected in their NIR spectral features.

\section{Observation}

The 2--2.5 $\mu$m spectra, studied in this paper, were obtained with the
CGS4 spectrometer (Mountain et al.  1990) on UKIRT on the nights 
of 1992 November 5 and of 1993 November 24, 25, and 26. 
A 75 l/mm grating was used in first order, providing a resolving
power ($\lambda/\delta\lambda$) of $\sim$300 in the $K$-band and a spectral
coverage encompassing most of the $K$-band.  The plate scale was 3 arc--second
per pixel and the slit width was 1 pixel.  This aperture was sufficient to cover
the central and most luminous part (usually the nucleus) of the galaxies.  The
slit length was 30 pixels.  For all galaxies, the slit was positioned in the
north south direction on the brightest optical peak as seen from the offset
guider camera.  Integration times were generally 10--20 minutes and sky frames
were obtained by nodding 10 pixels along the slit, hence keeping the object
always on the slit.  Several spectroscopic and photometric standards were also
observed during each night.  The spectrophotometric accuracy of our spectra is
estimated to be $\sim$20\%.  The spectra were reduced under the IRAF environment
using IRAF scripts written by one of us (RD).  Details on the procedure followed
for the reduction can be found in Doyon et al.  (1995).

In Table~3, we present the fluxes and equivalent widths of the Br$\gamma$ and
H$_2$ emission lines as measured in our spectra.  These quantities were estimated
after rectifying the continuum by fitting a power law
($F_\lambda\propto\lambda^{\beta}$) from the featureless sections of the
spectrum. The choice of a power--law is physically justified by the fact 
that late-stars show approximately Rayleigh--Jeans 
spectrum  in the NIR (F$_{\lambda} \propto \lambda^{-4}$, between 2 and 2.9$\mu$m; 
see appendix A of Doyon, Joseph \& Wright 1994). A $\chi^2$ minimisation routine was then used to fit a Gaussian
profile to the Br$\gamma$ emission line. In Table~3, values without an error estimate
corresponds to 3 sigma upper limits.

We give in Table~4 the exponent ($\beta$) of the power law fit for the
continuum, the CO spectroscopic indices (CO$_{spec}$), as
defined in Doyon, Joseph \& Wright (1994),
and the spectroscopic K magnitudes.  The CO$_{spec}$ quoted include the 
uncertainty in the continuum level. In general, the error on CO$_{spec}$ is dominated by this
uncertainty, which is less than ~0.02 mag. The second column
of Table~4 gives the redshifts as deduced from the features observed.  These redshifts
are in relatively good agreement with the one found in NED. The last column
gives the signal to noise ratios in our spectra.

For each galaxy in the two tables we give two sets of data. The first set
corresponds to values as measured using an aperture covering 3$\times9$ 
arc--second (Aperture 1). The second set (the last four columns of each 
tables) corresponds to values as measured using an aperture covering only 
3$\times3$ arc--second (Aperture 2). In our following analysis, the smaller
aperture will be associated to the nucleus of the galaxies.

\begin{figure*}
\hbox{
\centerline{\psfig{figure=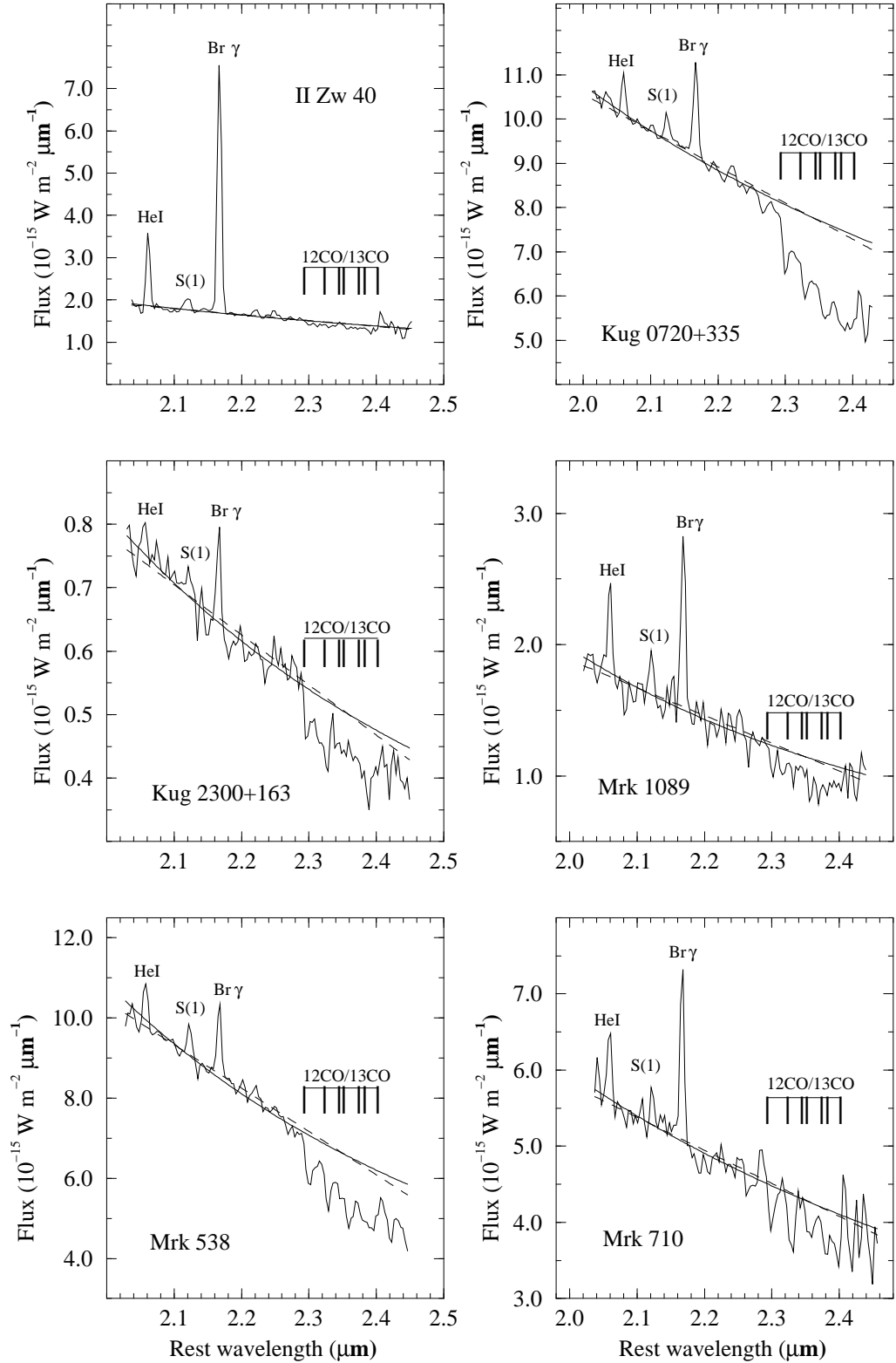,height=600pt}}}
  \caption{Some examples of spectra with strong Bracket
$\gamma$ emission--line.  The spectra were shifted to their rest frame using velocities in
Table~1.  Conspicuous emission lines and absorption features are identified.
For each galaxy we show the power law fitted on the continuum and the window
used to measure CO$_{spec}$. We also
show a straight line fit. A straight line do not give a good fit for
the K continuum of late--stars and galaxies. However, it would
have no significant effects on the measurements of 
the CO$_{spec}$ (variation $<1$\%) }
\end{figure*}

\begin{figure*}
\hbox{
\centerline{\psfig{figure=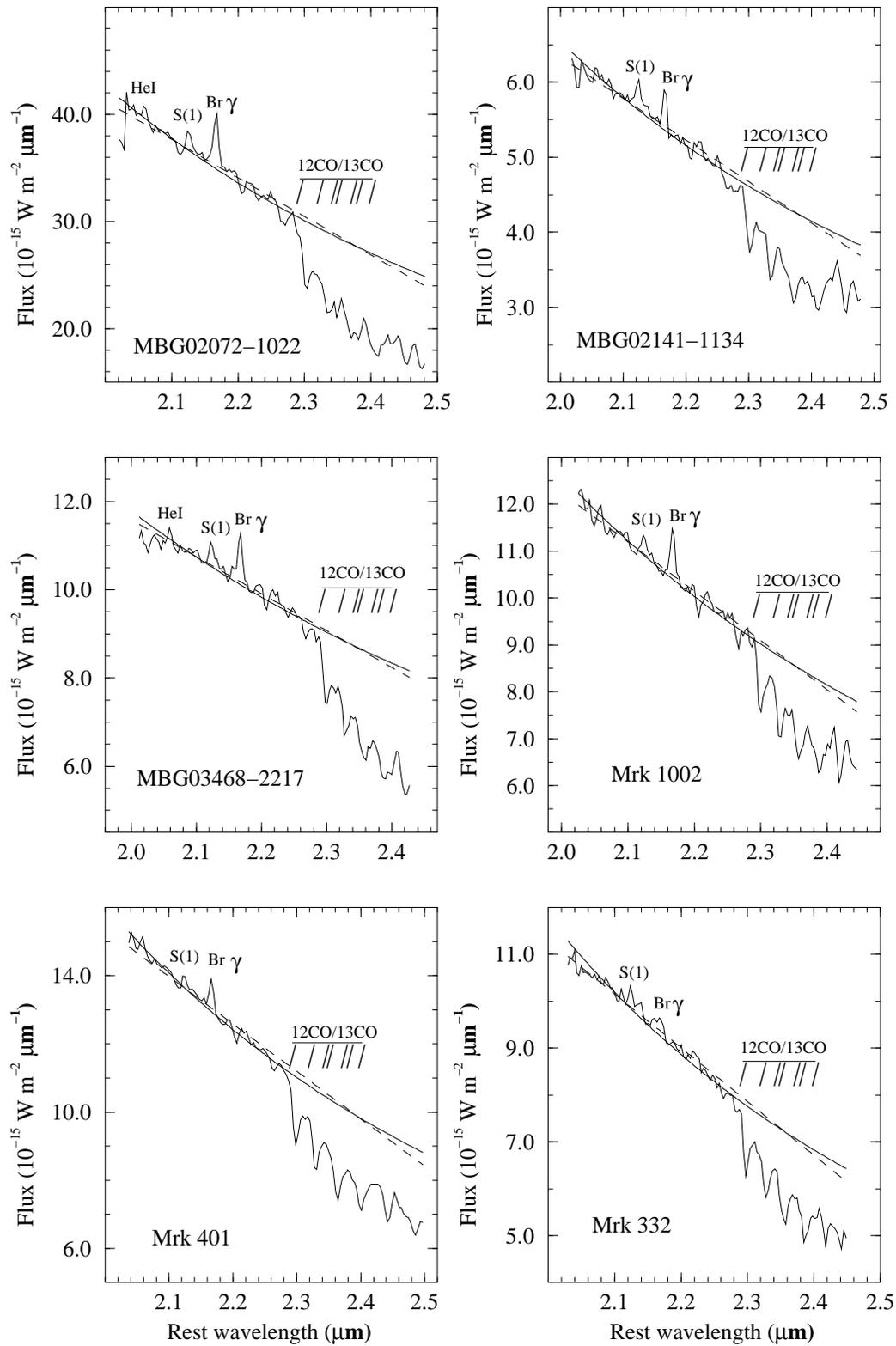,height=600pt}}}
  \caption{Examples of spectra with relatively weak Bracket~$\gamma$ 
emission--line and strong
CO lines.  The spectra are as described in Figure~2}
\end{figure*} 

A few representative spectra are shown in Figure~2 and Figure~3.  We regroup the spectra in
two categories.  In the first category, Figure~2, the galaxies possess strong
Bracket~$\gamma$ (2.166 $\mu$m) emission.  These galaxies also show
relatively strong He{\rm I} (2.059 $\mu$m) and H$_2$ 1-0 S(1) (2.122
$\mu$m) emission lines.  They show, on the other hand, different strengths of
the CO absorption band longward of 2.293 $\mu$m.  

The most extreme case in our sample 
is the H{\rm II} galaxy {\rm II}~Zw~40. This galaxy shows very strong emission lines and 
extremely weak CO band. This galaxy also happen to be one of the less chemically evolved 
galaxies in our sample. Two SBNGs with intense emission lines
are Mrk 710 and Mrk 1089. Most galaxies
in our sample resemble Kug 0720+335 and Mrk 538 (NGC~7714), which have
medium intensity emission lines and relatively strong CO band.  

In the second category, Figure~3, we find galaxies which 
have weaker emission lines than Mrk~538 and somewhat stronger 
CO band. In some galaxies, like Mrk~332, the NIR emission lines are
barely visible, and only upper limits can be measured.

The NIR spectra of our sample of galaxies show significant variations.
This result indicates that the diversity of physical
characteristics of starburst galaxy hosts, as observed in the optical 
and far infrared, implies some variety of the NIR spectral features.
Synthetic starburst models will now be used in order to understand
what these variations mean in terms of the
burst characteristics.

\section{NIR Analysis}

\subsection{Synthetic starburst models:
duration, age of the burst and IMF} 

Using synthetic starburst models, it is possible to interpret
NIR features like EW(Br$\gamma$) and CO$_{\rm
spec}$ in terms of the duration of the burst, its age and the form of the IMF
(Puxley, Doyon \& Ward 1997; Leitherer \& Heckman 1995; Mayya 1997).  
The principal steps in the synthetic starburst model are the following (Leitherer
\& Heckman 1995): A) stars are formed at specified rates and distributed along the
HR diagram; B) evolutionary models describe the time dependence of the individual
physical parameters of the stars (mass, luminosity, etc.); C) the stellar number
densities in the HR diagrams are determined and any desired synthetic quantities
assigned to each stars. By summing over the entire stellar
population the model yields the {\it integrated
properties} of starburst galaxy. 

In these models, star formation is represented by an exponential
law:  $\xi(t)=\xi(t=0) \exp{-t/\tau}$, where $t$\ is the age of the burst
and $\tau$\ its duration.  Two limiting cases are of special interest: an
instantaneous burst and a constant star formation rate.  An instantaneous
burst corresponds to the case where the duration of star formation is
short as compared to the age of the galaxy ($\tau << t$). All the stars are 
thus formed at the same epoch.
The other limiting case is a constant star formation rate (CSFR).
This mode of star formation is characteristic today of disc in giant spiral
galaxies (Kennicutt 1983).

It is important to note that theoretically a
CSFR can always be approximated as a series of instantaneous bursts. 
It means that a system having experienced a sequence of
bursts over a time period comparable to $\tau = t$ would be observationally
indistinguishable from a galaxy with a CSFR (Leitherer et. all 1999).

Another important parameter in synthetic starburst model is the IMF.
This function is usually expressed as a power law:  $\phi(M) =
Cm^{-\alpha}$, where the constant $C$\ is determined by the total mass
converted into stars.  To get the whole spectrum of star masses formed, the
IMF is integrated between the upper and lower mass cutoff, M$_{up}$ and
M$_{low}$.  In starburst models the exponent of the IMF and the two mass
cutoff are not very well constrained and it is
therefore important to try different possible values.

We choose to compare our data with the models of Leitherer et al. (1999).
These authors have made their grids of results directly
accessible on the WEB\footnote{http://www.stsci.edu/science/starburst99}.
We consider only two limiting cases:
the instantaneous burst and the CSFR (1 M$_\odot$ per yr).  For each of these
cases, three different scenarios are tested: 1) a Salpeter IMF, $\alpha =
2.5$, with M$_{up} = 100$\ M$_\odot$; 2) a Salpeter's IMF with M$_{up} = 30$\
M$_\odot$; 3) an IMF with a steeper slope, $\alpha = 3.3$ and M$_{up} =
100$\ M$_\odot$.  The models cover ages from $10^6$ to $10^9$ yr.

Comparisons of our data for EW(Br$\gamma$) and CO$_{spec}$ with the
instantaneous burst models are presented in Figure~4. From this figure, we
conclude that none of the instantaneous burst models can fit the ensemble of our data. 
For solar metallicity (Z\ $=0.02$) the models generally underestimate the value
of EW(Br$\gamma$). Some galaxies can be fitted by instantaneous bursts, but the predicted 
metallicity is 2 times solar, which is much higher than the gas metallicity we have
estimated in these galaxies. Our results confirm those of Goldader et al. (1997b),
extending their conclusion to starburst galaxies with different physical properties and FIR luminosity.
It implies that the instantaneous burst scenario is not a good
representation of the starburst phenomenon.

\begin{figure}
\hbox{
\centerline{\psfig{figure=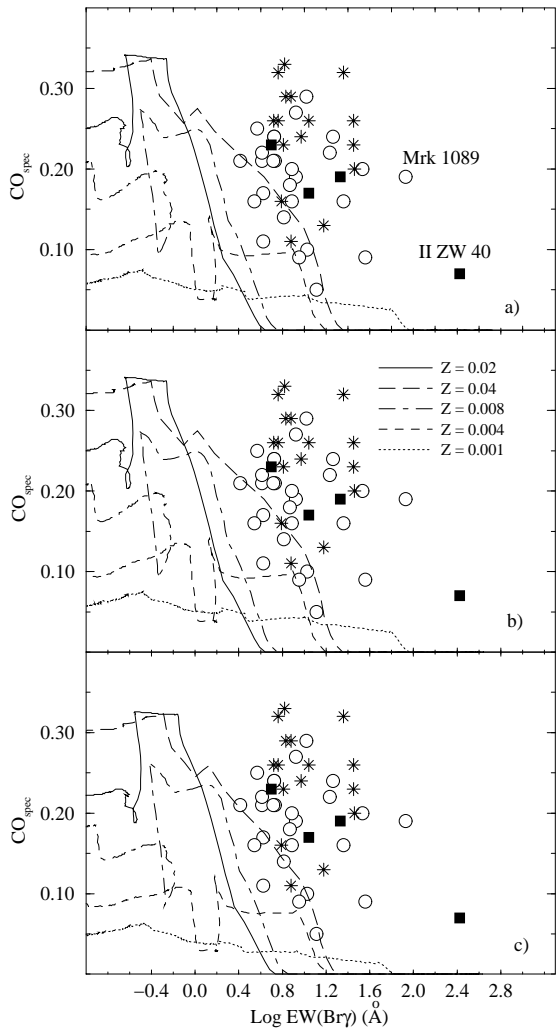,height=400pt}}}
  \caption{Comparison of our data with instantaneous burst models.
a) IMF with slope $\alpha = 2.5$ and M$_{up} = 100$\ M$_\odot$; 
b) $\alpha = 2.5$ and M$_{up} = 30$\ M$_\odot$;
c) $\alpha = 3.3$ and M$_{up} = 100$\ M$_\odot$.  The models
cover ages from $10^6$ to $10^9$ yr. Open circle = SBNG, filled square = H{\rm II} galaxy, 
star = LIRG}
\end{figure}

Our data are compared with the CSFR models in Figure~5.
One can see in Figure~5a that the Salpeter's IMF model with M$_{up} = 100$\ M$_\odot$
cannot fit the ensemble of our data.   
But a relatively good fit is obtained in Figure~5b, when the upper
mass cutoff is equal to M$_{up} = 30$\ M$_\odot$ . We see  
no improvement in Figure~5c in adopting an IMF slope $\alpha = 3.3$.  

\begin{figure}
\hbox{
\centerline{\psfig{figure=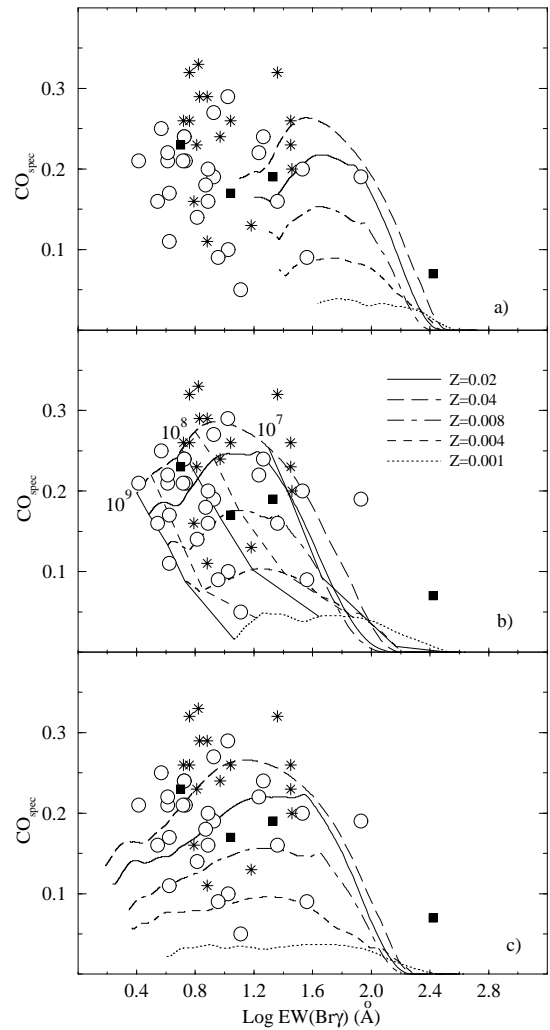,height=400pt}}}
  \caption{Comparison of our data with constant star formation models (1 M$_\odot$ per yr).
a) IMF with slope $\alpha = 2.5$ and M$_{up} = 100$\ M$_\odot$; b) $\alpha = 2.5$ and M$_{up} =
30$\ M$_\odot$; c) $\alpha = 3.3$ and M$_{up} = 100$\
M$_\odot$. The signification of the symbols are the same as in Figure~4}
\end{figure}

Once again our results confirm those of Goldader et al. (1997b), extending their conclusion
to all the starburst types known. We conclude that the best parameters
which describe starburst galaxies in the nearby Universe are a CSFR and a Salpeter's
IMF with an upper mass cutoff M$_{up} = 30$\ M$_\odot$.

According to this solution, the bursts have ages
between 10 Myr and 1 Gyr, the majority being older than 100 Myr. 
Comparing with LIRGs, we found, on average, younger bursts for these galaxies, 
with ages lower than 100 Myr. This is consistent with the difference in
luminosity ratios noted in Section~2.5. The model reproduce correctly also
the fact that LIRG are more metal rich than the SBNGs. The consistency with
the metallicities deduced from their optical spectra (Section 2.2) is remarkably good.

According to the model, H{\rm II} galaxies must have younger bursts than the SBNGs.
The metallicities predicted by the models are also in good agreement with those
measured. Young bursts and low metallicities are also consistent
with the higher dust temperatures suggested by the FIR model in Section 2.3.

In our sample, the galaxy which have the youngest burst is
{\rm II}~Zw~40. The position of this galaxy in Figure~5a suggests that it may be
possible to fit an IMF with upper mass cutoff M$_{up} = 100$\ M$_\odot$ (although the 
predicted metallicity would then be wrong).
The same conclusion may also apply to Mrk 1089.
These observations may suggest varying IMF 
in some galaxies (Doyon, Puxley \& Joseph 1992).

From the above discussion, we conclude that the solution
of the synthetic starburst model is consistent with observations in
the optical and FIR. In particular, it satisfactorily explains the differences
between the different starburst types: H{\rm II} galaxies have younger bursts and
lower metallicities than SBNGs, while LIRGs have younger bursts but higher metallicities.
These differences may be related to different histories of star formation or 
different stages in burst evolution.

Taken at face value, the above result confirms the studies
of Doyon, Puxley \& Joseph (1992), Goldader et al. (1997b) and
Coziol (1996). The generality of our observation further suggests that this must be
a characteristic of starburst galaxies in the nearby Universe. However, the above solution
may not represent strong constraints on the duration of the burst or the IMF.
This is because the synthetic starburst models we used cannot distinguished
between real constant star formation and a series of instantaneous bursts
(Leitherer et. all 1999). In order to verify
this alternative, we will now reexamine our spectra
and search for traces of previous bursts.

\subsection{Spatial variation of NIR features: evidences for sequences
of bursts}

\begin{figure}[b]
\hbox{
\centerline{\psfig{figure=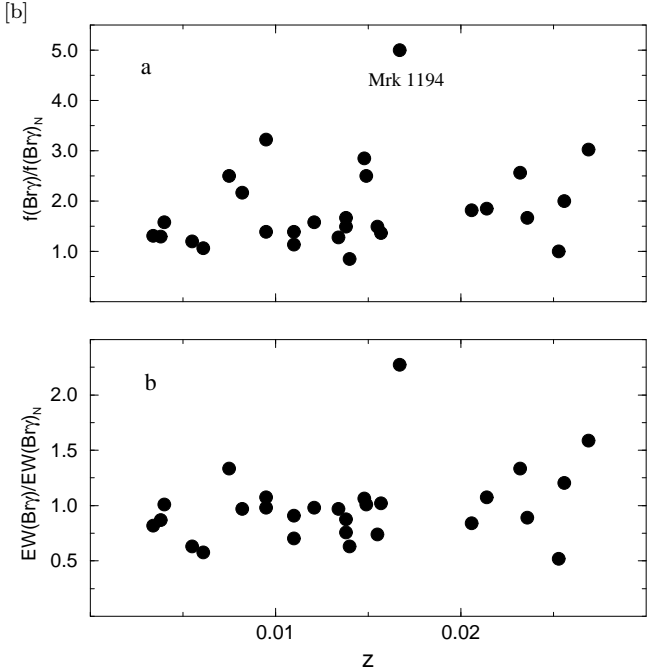,height=250pt}}}
  \caption{Comparison of the ratios of Br$\gamma$ fluxes (a) 
and of of the equivalent widths (b), as measured in and out 
of the nucleus and as function of the redshift.
In this analysis we associate the smaller aperture 
($3\times3$ arc--second) to the nucleus.
No relation with redshift is observed, which indicates that
the increase of scale with distance, implied by 
our two fixed apertures, does not introduce any systematic variations.
The galaxy showing the higher variation is Mrk 1194. No particular reasons
explain this variation }
\end{figure}

In order to reproduce CSFR using a sequence of bursts, one must assume that
this sequence spreads over a relatively long period of time. The 
solution of synthetic starburst model suggests a period extending over $10^8$ up to
1 Gyr. If this sequence of burst is related to some self--regulated mechanism or
to propagation of star formation, we may also assume that the different bursts
do not happen in the same regions of space. This implies that the bursts regions should
be heterogeneous in space and in time. With sufficient spatial resolution, 
we should then observe a variation in
burst characteristics as we survey different regions.
As a first test, therefore, we can check if a variation of
aperture in our spectra induces some variation of the NIR spectral features.

Before doing the aperture test, it is important to verify
that the obvious increase of scale with distance, implied by 
our two fixed apertures, do not introduce any systematic variations, which
could be misinterpreted as variation of burst characteristics.
Associating the smaller aperture ($3\times3$ arc--second) to the nucleus,
we verify in Figure~6 that the ratios of the Br$\gamma$ fluxes
and equivalent widths, as measured in and out of the nucleus, do not show
any relation with the redshift.
Similar negative results were also obtained for H$_2$, the K magnitudes, the
CO$_{spec}$ and for the slope of the continuum ($\beta$). These negative results suggest
that what ever variations we may observed for these ratios, their origin must
be intrinsic to each galaxy.

Having verified that the two fixed apertures do not introduce
spurious effects, we now proceed to examine how 
the different ratios varied in each galaxy.
In Figure~7a, we first compare the ratios of the fluxes in Br$\gamma$ 
with the ratios of the equivalent widths.    
In general, the ratios f(Br$\gamma$)/f(Br$\gamma$)$_N$ are higher than one, which
indicate that the bursts of star formation generally extend over a few kpc around
the nucleus.
The equivalent widths, on the other hand, are
generally wider in the nucleus, except in galaxies where the
ratio f(Br$\gamma$)/f(Br$\gamma$)$_N > 2$, which correspond to 
extended bursts, where the equivalent width stays constant
or increases out of the nucleus. 
According to synthetic starburst models, EW(Br$\gamma$)
is an indicator of the age of the burst.
From the above observation we then deduce that in the galaxies where
the star formation is more concentrated, the bursts are younger in the nucleus.
The contrary seems to be true in galaxies where the bursts are more extended.
No difference is observed between H{\rm II} galaxies and SBNGs.

\begin{figure*}
\hbox{
\centerline{\psfig{figure=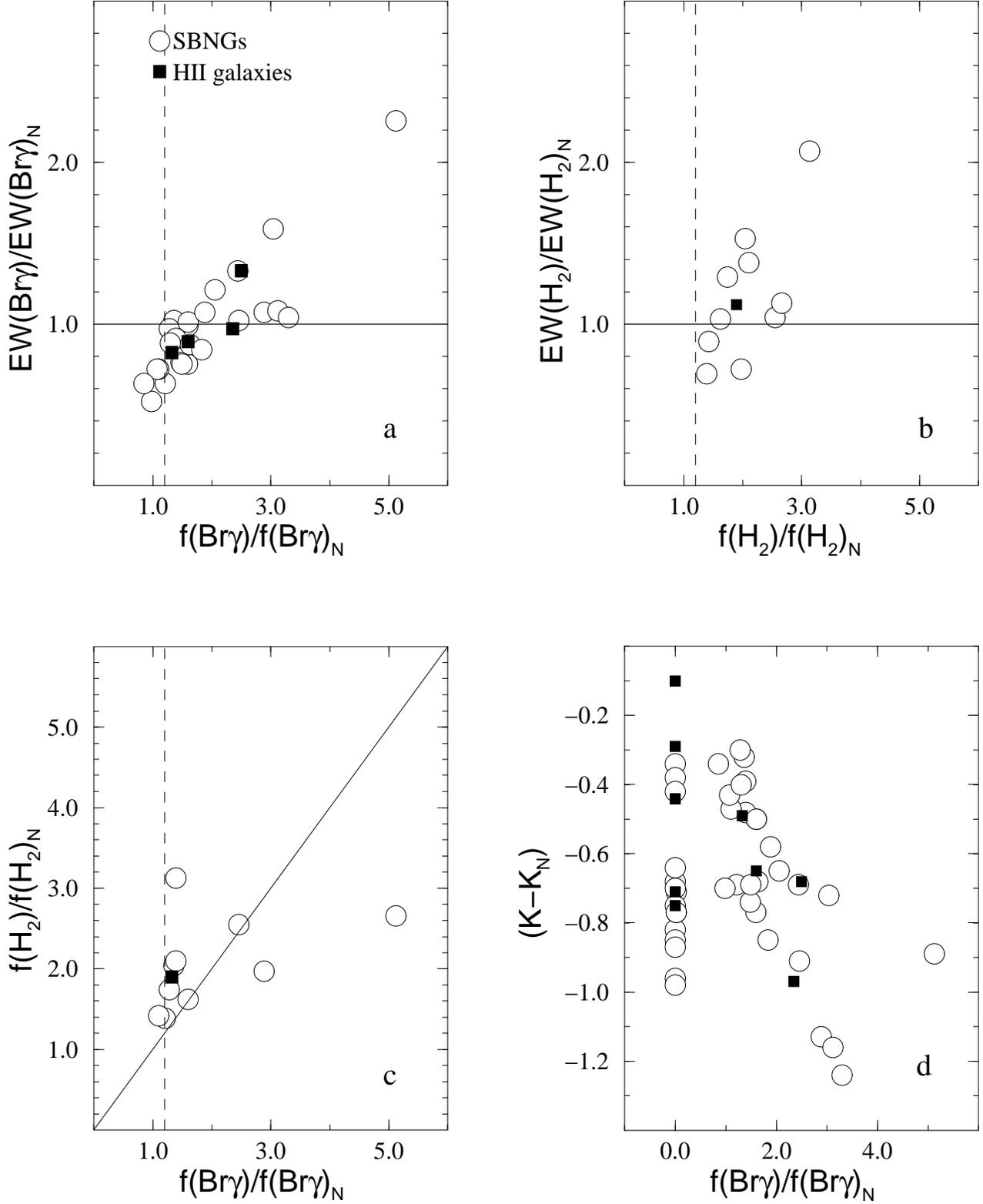,height=600pt}}}
  \caption{Variation of the NIR features as measured in two different
apertures.  We compare the values measured in a 3$\times$3 arc--second aperture
(nucleus) with those in a larger 3$\times$9 arc--second aperture.  
The vertical dashed line in a, b and c is the limit adopted for spatial resolution}
\end{figure*}

The case for H$_2$, Figure~7b, is slightly different. 
Very few galaxies in our sample show significant H$_2$ emission.
In these galaxies we find the same tendency than for Br$\gamma$, which is the
H$_2$ emission generally spreads out of the nucleus. But 
the H$_2$ equivalent width tends to be wider out of the nucleus. 
In Figure~7c one can see that in galaxies where the star
formation is concentrated in the nucleus (f(Br$\gamma$)/f(Br$\gamma$)$_N < 2$),
the H$_2$ emission is more intense than Br$\gamma$ out of the nucleus. 
Only two galaxies, Mrk 1089 and 1194, show the reverse tendency.
For the galaxies in our sample with significant amount of H$_2$ emission, 
the H$_2$ emission seems, therefore, to be always more intense and wide spread
in regions where Br$\gamma$ is less intense and less abundant. We verified also that
50\% of the galaxies without H$_2$ detection have detectable Br$\gamma$ emission,
which is consistent with the general tendency observed. The signification of
this variation is not obvious, however, and its possible
interpretation will therefore be kept for later during our analysis.

For the K--band magnitude, we naturally expect it to increase as
the number of stars included in the aperture increases. However,
the mean variation observed in Figure~7d seem smaller than what we expected.
While the gain in surface represented by the difference in aperture is a factor 3,
we find a meager $\Delta$K $=-0.7$, which corresponds to an increase in flux by
only a factor 1.8. The K--band emission, therefore, seems mostly concentrated in the nucleus.
Indeed, one can see in Figure~7d that the difference in K--band magnitude
is smaller in galaxies where the bursts are concentrated (f(Br$\gamma$)/f(Br$\gamma$)$_N
<2$) than in galaxies where the bursts are extended. The K--band emission seems, therefore, to
tightly follow the region of the bursts. This observation suggests
that the stars emitting in the K band must be tightly related to the bursts.

No trend was observed for CO$_{spec}$ or $\beta$. 
These two parameters show random variations with standard deviations $\pm 0.04$ for
the difference in CO$_{spec}$ and $\pm 0.15$ for the ratios of $\beta$.

To illustrate what effects the variations of NIR features have on
the burst characteristics of each galaxy, we show in Figure~8
the changes introduced by an increase in aperture in the burst model of Figure~5b.
In this figure, an increase of the equivalent width transforms into a younger age while
an increase of the CO$_{spec}$ transforms into an increase of metallicity and
a younger age. In Figure~8, one can see that some variations may imply significant
differences in terms of the burst characteristics. No trend is observed, however,
which suggests that the cause of these variations may depend on the
particular history of star formation of each galaxy. Higher
spatial resolution will be necessary to confirm this result.
 
From the above analysis we conclude that the variation of the NIR spectral features
with aperture are consistent with a variation of burst characteristics in space.
This result supports our hypothesis that the bursts regions are not homogeneous in space.
However, the absence of systematic trends and the lack of spatial resolution impede
us to conclude on the real cause of this variation and we must therefore search
for other evidence in favour of the multiple burst hypothesis.

\begin{figure}
\hbox{
\centerline{\psfig{figure=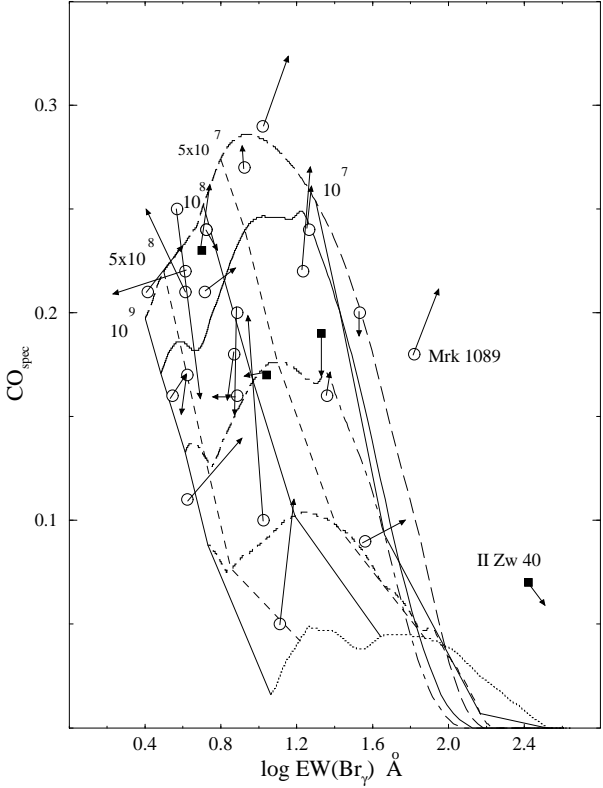,height=300pt}}}
  \caption{Variation of the burst characteristics with 
a variation in aperture. The parameters of the model are the same as in Figure~5b}
\end{figure}

\subsection{Luminosity--luminosity relations}

\begin{figure}
\hbox{
\centerline{\psfig{figure=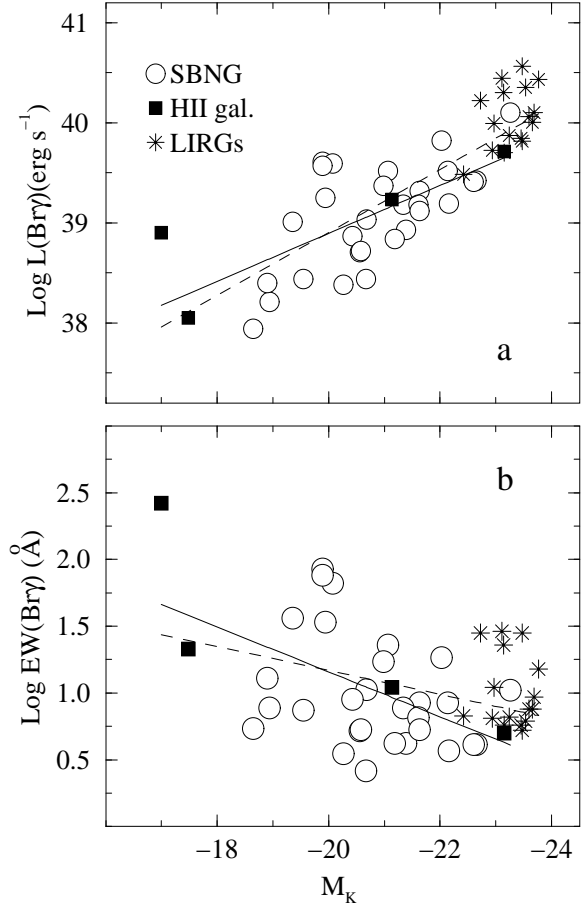,height=350pt}}}
  \caption{Relation between Br$\gamma$ and absolute K 
magnitude. The continuous curves are linear regressions 
fitted using only our galaxies. The dash curves are linear regression
fitted adding the LIRGs to our galaxies. The coefficients of correlation
are given in the text}
\end{figure}

\begin{figure}
\hbox{
\centerline{\psfig{figure=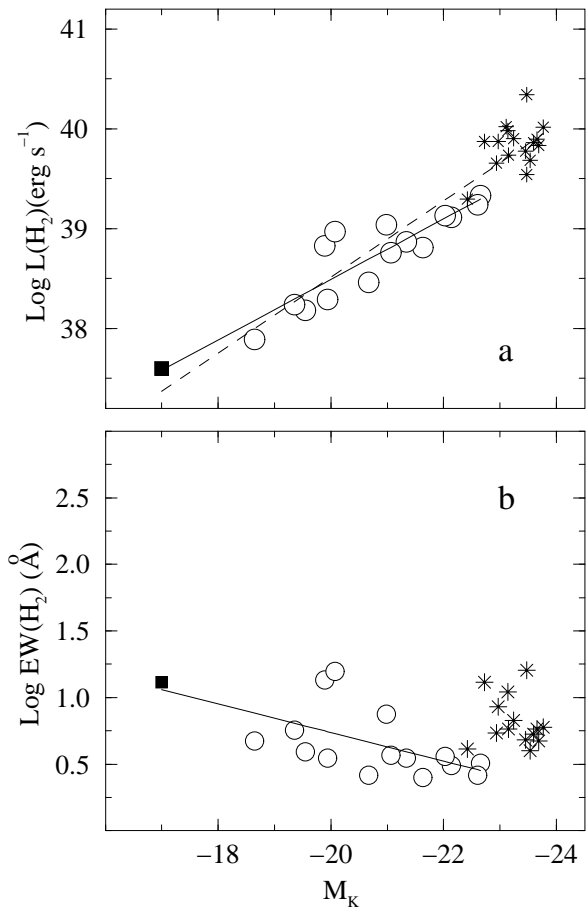,height=350pt}}}
  \caption{Relation between H$_2$ and K--band absolute 
magnitude. The signification of the symbols and
curves are as in Figure~9}
\end{figure}

\begin{figure}
\hbox{
\centerline{\psfig{figure=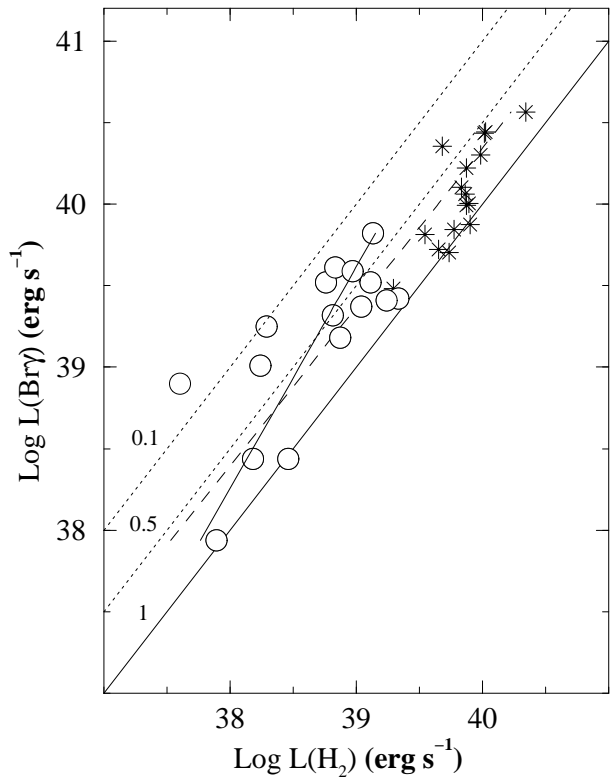,height=300pt}}}
  \caption{Relations between Br$\gamma$ and 
H$_2$ luminosities.  The signification of the symbols are as in Figure~9.  
The diagonals correspond to ratios 1.0, 0.5 and 0.1, consistent with values for shocks or
fluorescence.  The short continuous curve is a linear regression on our
galaxies.  The dash curve is a linear regression including LIRGs }
\end{figure}

\begin{figure}
\hbox{
\centerline{\psfig{figure=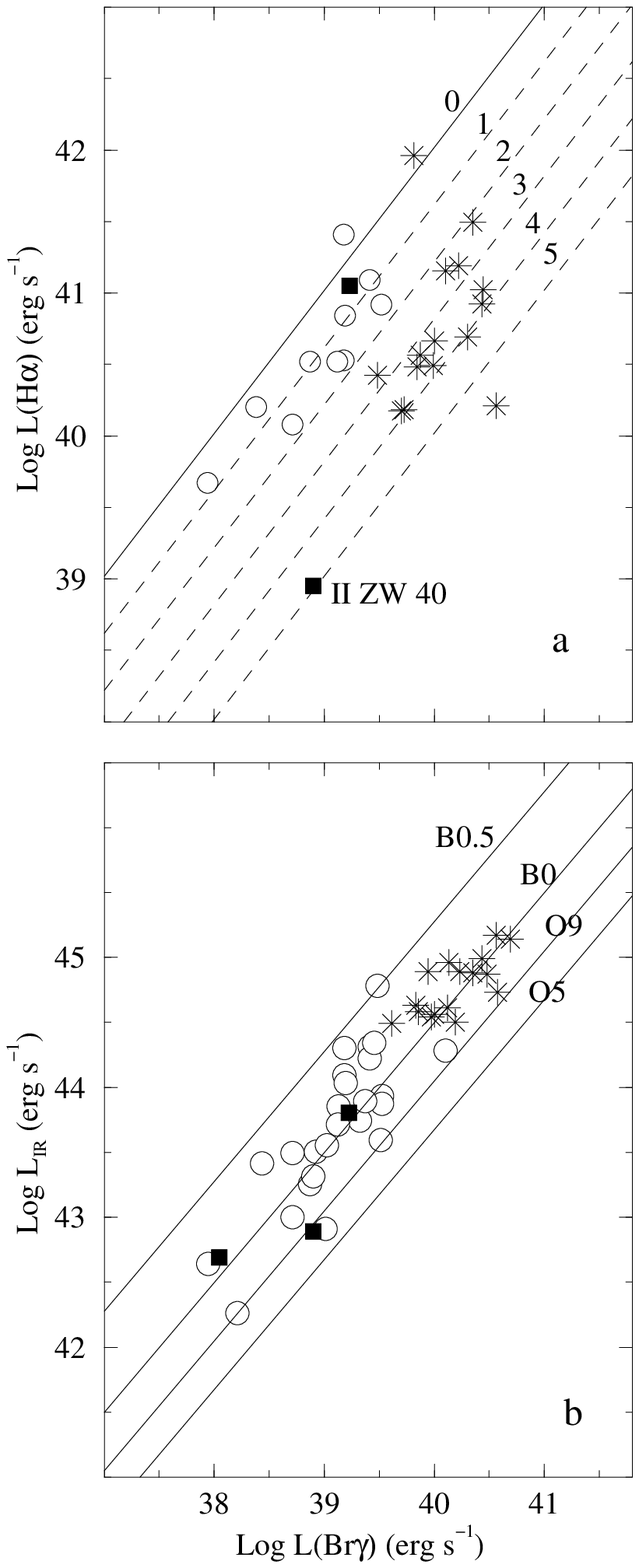,height=500pt}}}
  \caption{a) Relation between H$\alpha$ and Br$\gamma$ luminosities.
The different diagonals
correspond to theoretical ratios between the luminosities, assuming various
magnitudes of extinction as indicated by the numbers; b) Relation
between the FIR and Br$\gamma$ luminosities. The diagonals are
theoretical ratios expected for different spectra types.  The luminosities of
Br$\gamma$ were corrected for extinction as determined in a}
\end{figure}

Our previous analysis suggests that the stars emitting in
the K--band are tightly related to
the burst population. We should then expect to find a
good correlation between the K--band absolute magnitude 
and parameters link to the burst. This is important to verify,
because the stars producing the K--band luminosity are probably  
the same stars which produce the CO band. A correlation of K--band luminosity
with parameters directly linked to the burst would thus favour massive and young stars, namely
RSGs, over older stars like RGs.

We see in Figure~9a that L(Br$\gamma$) increases with the K--band absolute magnitude.
A linear regression performed on our data (continuous curve in
Figure~9a) gives a weak but statistically significant coefficient of correlation R$=67$\%.
(with 33 points, the probability to obtain higher correlation by chance
is P$_{33,0.67} < 0.05$\%; Pugh \& Winslow 1966).
The correlation improves to 82\% when we include the
LIRGs (dash curve in Figure~9a). The properties of our galaxies are in good
continuity with those of the LIRGs, which form the upper limit of the distribution.

The correlation of L(Br$\gamma$) with the K--band absolute magnitude confirms
that the stars producing this luminosity are related to the
burst population. The relative weakness of this relation, however,
may suggest that they are not correlated to the younger stars.
But this result is consistent with what we expect if the stars
emitting in the K--band are in majority RSGs:
the number of RSGs increases with the intensity
of the burst, but these stars also appear in great number
only after the peak intensity of the bursts.

In Figure~9b, we find a reverse tendency for EW(Br$\gamma$),
this parameter decreasing as M$_K$ increases. These two parameters 
are not strongly correlated (R$=57$\%), although the correlation
is still statistically significant (P$_{33,0.57} < 0.05$\%).
The obvious interpretation for this reverse tendency is
that galaxies which are richer in burst regions naturally also contain the highest number
of evolved stars.

What is interesting in this last figure is the high dispersion of the data.
This high dispersion suggests that the evolution of the bursts depends
on the particular star formation history of each galaxy. This is consistent with
our analysis of the variation with aperture and suggests
that the bursts may also be heterogeneous in time (the dispersion
in EW(Br$\gamma$) is a dispersion in time). The behaviour
of the LIRGs is particularly remarkable, suggesting younger bursts than the observed
tendency would have predicted. This could be one indication of repetitive bursts.

Note that we do not reject, a priori, the alternative which is
that the stars emitting in the K--band are old RGs. In this case, 
the correlation of M$_K$ with L(Br$\gamma$) would suggest some relation between
mass and the intensity of the bursts, while the correlation between M$_K$ and EW(Br$\gamma$)
would mean that the bursts evolve more rapidly in galaxies where they are more intense.
However, it would be difficult in this case to understand why the correlation are not stronger.
But such interpretations, in fact, are much more constraining than what
our data can revealed (we do not have information on the masses
or the intensity of the bursts, which, anyway, should be normalized by
the surface or volume). To our knowledge, such relations were
never reported before, and our data are obviously insufficient to defend
them. Our favoured interpretation, on the other hand, do not
imply any new mechanisms, relying only on standard star evolutionary scenarios.
Another argument in favour of our interpretation may be the relatively strong correlation
we find between L(H$_2$) and M$_{\rm K}$, as we now show.

A very strong correlation, R$=92$\%, is found between
L(H$_2$) and M$_{\rm K}$ in Figure~10a. The lower number of
points in this fit do not influence the results:
with 16 points, the probability to obtain a higher correlation by
chance is P$_{16,0.92} < 0.05$\%. This good correlation do not depends
on the inclusion of the LIRGs either, the coefficient increasing only
to R$=94$\% when we include them. For the equivalent widths, Figure~10b,
we also find a better correlation than for Br$\gamma$, R$=62$\%, P$_{16,0.62} < 1.4$\%.
This behaviour suggests that the H$_2$ emission is better
correlated to a more evolved phase of the bursts, being
correlated to SRGs, than to a younger phase.
In general, therefore, the behavior of H$_2$ emission seems consistent with the
hypothesis of heterogeneous burst in space and time, showing
variation with aperture (section 4.2)
and the coexistence of evolved components with younger ones.
It would help, however, to determine what is the process
producing this line, since the two most probable
mechanisms, supernovae shocks remnants and UV-fluorescence,
imply much different time scales.

In principle, it would be possible to determine what is
the mechanism producing the H$_2$ emission by comparing the
two luminosities L(Br$\gamma$) and L(H$_2$).
In Figure~11 we find that L(H$_2$) is relatively well correlated to
L(Br$\gamma$) (R$=74$\% with P$_{16,0.62} < 0.4$\%).
But the ratios L(Br$\gamma$)/L(H$_2$)
stay between 0.1 and 1, which is consistent with supernovae
shocks remnants (Moorwood \& Oliva 1990) or UV-fluorescence
(Puxley, Hawarden \& Mountain 1990). Therefore, although our observations
suggest that the H$_2$ emission phenomenon is related to
a more evolved phase of the bursts, we are unable to say if this phase
happen a few million years (the time scale of supernovae) or a
few $10^7$ or $10^8$ yrs (the time scale of B stars, responsible for UV-fluorescence)
after the maximum of the bursts.

\subsection{Spectral types of ionising stars}

In Coziol \& Demers 1995 and in Coziol (1996), one evidence
in favor of a sequence of bursts was the predominance of
B type stars in starburst nucleus galaxies. We can repeat
the analysis performed in Coziol (1996) to verify this result in the NIR.
The advantage of working in the NIR is that the
extinction in this part of the spectrum is less severe, and the fraction of 
obscured stars should therefore be less important (Calzetti et al. 1995), decreasing the 
uncertainties related to this method. 

The level of extinction in our galaxies can be estimated (Figure~12a) by comparing
L(Br$\gamma$) with the luminosity in H$\alpha$ (L(H$\alpha$)).
For the theoretical relation between L(Br$\gamma$)
and L(H$\alpha$) we use the one adopted by Leitherer \& Heckman (1995).
The theoretical ratio is traced in Figure~12a as a continuous curve.
Lower ratios then must correspond to various
magnitudes in extinction in the optical. 
In our galaxies, the extinction varied between 0 and 2 magnitudes , which is
in good agreement with values estimated from the Balmer
decrement method (Contini, Consid\`ere \& Davoust 1998). This should not be surprising
considering the low dust extinction in these galaxies. The LIRGs seem to suffer 
slightly higher extinction, the values varying between 3 and 4 magnitudes. 

Once corrected for extinction, the flux ratios Br$\gamma$/L$_{\rm IR}$ are proportional
to the ionising flux produced by stars with various spectral types (Devereux \& Young 1990).
We show in Figure~12b the locus expected for ionising star clusters
with different spectral types. This figure confirms
the dominance of B type stars as found in Coziol (1996).

\section{Discussion}

The question that we need to answer is: can we accept
the result of the synthetic starburst model? In order to answer this question we need
first to better understand this result. The main reason why the instantaneous
burst solution is rejected by our observation is because the predictions of the
models are in contradiction with the values of the CO spectral indices we measured. 
In an instantaneous burst, the Br$\gamma$ equivalent width stays high for the first $10^7$ years. 
During this time, CO$_{spec}$ is negligible and begins
to increase only around $10^7$ year, when the most massive stars evolve as
RSGs. It is because we observe both intense Br$\gamma$
emission and relatively strong CO lines that we need continuous star formation.
The answer seems, therefore, to rely on the validity of the CO spectral index.
 
To obtain a good fit with any of the instantaneous burst models, the CO spectral indices
would need to decrease by more than 50\%. Obviously, this cannot be due to the method
we used to measure these lines. The strength of the CO bands in our spectra is undeniable.
A relatively strong value of CO$_{spec}$ is consistent with a predominance
of RSGs over RGs (Doyon, Joseph \& Wright 1994).
This is consistent with the correlation found between M$_K$ and L(Br$\gamma$),
which suggests that the stars producing this luminosity, and probably also the CO band,
are related to the burst population. Moreover, this relation is relatively weak because
RSGs appear in great number only slightly after the culmination of the burst.
The relatively strong CO bands in our spectra seems therefore
to be real, and a constant star formation rate is consequently necessary
to explain our data.

But what about the low upper mass cutoff? If we look at Figure~5, we see that the
reason why we need such feature is because the Br$\gamma$ equivalent widths 
seem too weak as compared to the model with upper mass cutoff M$_{up} = 100$\ M$_{\odot}$.
Now, this result is far from obvious. It is therefore highly significant that when
we determine the spectral types of the ionising stars in these galaxies we
find a value which is consistent with this result: the dominant ionising stars
are early--type B or late--type O which have masses of the order
of 10 to 20 M$_\odot$ (Panagia 1973),
in good agreement with the upper mass cutoff suggested by the model.
We have therefore to conclude that the solution suggested by the synthetic
starburst model is fully consistent with our data.

But how could star formation be continuous?
In order to sustain star formation in starburst galaxies, some
mechanism other than interaction is necessary. Observations of nearby starburst galaxies
suggests that most ($\sim 70$\%) of these galaxies are relatively isolated and cannot be
associated to on--going interaction events (Telles \& Terlevich 1995; Coziol et al. 1997b).
In our sample, only five galaxies have a peculiar
morphology suggesting some sort of direct interaction. The best example
is NGC~1089. The galaxy {\rm II}~Zw~40, on the other hand, which also has the
youngest burst, cannot be easily connected to an interaction event.
The same is true for most of the galaxies in our sample.
More than two thirds (19) of our galaxies are
isolated early--type spirals (earlier than
Sbc) or galaxies with a compact appearance (14 galaxies).
The rest (10 galaxies) looks like normal late--type spirals (Sbc or later).
There is, consequently, no obvious dynamical cause for the bursts
observed in most of these galaxies, which suggests that some ''internal`` regulating
or self--sustained star formation mechanism must be active.

In order for star formation to be continuous over a significantly long period
of time, star formation rates cannot be very high either. This may be
in contradiction with some observations in very intense starburst galaxies.
This is why a sequence of bursts may be a more interesting alternative. If the
bursts are intense but do not last very long, 
the reservoir of gas will not be exhaust, allowing ignition of new bursts later on.
Assuming some self-regulated mechanism, this process could probably extend over
a relatively long period of time. One can already find in the literature different mechanisms
capable of producing bursts in sequence, either by regulation or propagation of star
formation from supernovae feedback (Gerola, Seiden \& Schulman 1980; Kr\"{u}gel \&
Tutukov 1993) or multiple small merger events on a few Gyr period (Tinsley \& Larson 1979).
These two alternatives were already found to be consistent with the chemical properties of
nearby starburst galaxies (see Coziol et al. 1997 and 1999 for example).

But what evidence do we have in favour of a sequence of bursts?
In our spectra we have found some variation of burst characteristics with the aperture
which suggests the bursts are not homogeneous in space.
The fact that we do not observe systematic trends for these variations
suggests that their origin may be related to the particular star formation history of
each galaxy. This could be one signature of internal processes.
Considering that these variations are observed 
over regions a few kpc wide, it is probably more realistic to assume also
that different star formation events occurred during different time intervals
(this would be the case for star formation propagation).
The dispersion of the K--band absolute magnitude vs. EW(Br$\gamma$) relation
is highly suggestive of some intermittent events, superimposed, perhaps, on a
more global behaviour.

However, the strongest evidence in favour of a sequence of bursts may be
the predominance of early--type B stars in the ionised regions of these galaxies.
Such stars, with masses of the order of 10 to 20 M$_\odot$, have lifetime of a few $10^7$ yrs only.
For one burst population, it is therefore difficult to understand how this
particular composition of stars is possible.
This cannot be explained solely in terms of an age effect. Indeed, the main sequence lifetime 
of these stars is only a factor 10 older than those of more massive stars and
a factor 10 younger than the age of the bursts deduced from the synthetic starburst model,
which would mean that we observe all these galaxies in a very particular 
and short phase of their evolution. Considering the generality of our
observations, this makes it a very improbable event\footnote{It is
doubtful that we can explain such rare events based on the obvious bias
that these are starburst galaxies, considering that such galaxies form between
25 to 30\% of all the galaxies in the nearby Universe.}.

A sequence of short bursts, on the other hand, without any limit on the upper mass cutoff
and extending over a relatively long period of time (a few 10$^8$ or 1 Gyr) in order to mimic continuous star formation, may reproduce such phenomenon. Short bursts would produce
few massive stars, which would then evolve
rapidly and all disappear before another burst begin. Depending on
the frequency of the bursts, the average population would then evolve statistically
towards the predominance of early--type B stars. Note that it is probably impossible to
reproduce such stellar population with constant star formation, unless the IMF upper mass
cutoff is truncated towards low mass stars.

To finish, we have concluded that we cannot identify what is the
mechanism responsible for the H$_2$ emission in our galaxies. However, the
different time scales over which these processes culminate should help us
estimate which is the more probable one. Supernovae shocks remnants
are predominant only when a high number of massive stars
are produced and the bursts are very young (a few $10^4$ years).
Infrared fluorescence, on the other hand, is produced by early B stars, which radiate ample photons
in the Lyman--Werner band ($\lambda = 912-1108$ \AA). These stars have longer life times and should therefore
prevail only when the burst is old or, for some reason, they are predominant in a
burst population. The fact that B stars do seem to be over abundant
in our galaxies suggests that fluorescence must be the main mechanism
responsible for the H$_2$ emission. One would then expect to see an increase of fluorescent emission
in regions where B stars dominate over O stars (Tanaka, Hasegawa \& Gatley 1991). This
would explain some of the variation we observed when we change
the aperture, which would constitute, consequently, another evidence in favour
of the multiple bursts scenario hypothesis.

\section{Conclusion}

Although the solution proposed
by the application of the synthetic starburst models
to our sample of starburst galaxies may look
counter intuitive, we have shown that it seems to be fully consistent with our data.
Our analysis confirms, therefore, the results previously obtained by
Deutsch \& Willner 1986, Doyon, Puxley \& Joseph (1992), Coziol \& Demers 1995,
Coziol (1996) and Goldader et al. (1997b).

However, the solution proposed by the models may not represent strong constraints on
the duration of the burst or the form of the IMF. This is  because
the models cannot distinguish between constant star
formation or a sequence of short bursts over an extended period of time.
This alternative, the sequence of short bursts, will have 
to be tested fully before we can conclude on the values of these
two important parameters in starburst galaxies.

What ever is the solution, our observations are clear on one thing, which is
that starburst must be a sustained or self--sustained phenomenon: 
either star formation is
continuous in time or different bursts happen in sequence
over a relatively long period of time. The generality of our observations 
suggests that this is a characteristic of starburst galaxies
in the nearby Universe.

\section*{Acknowledgments}

R. C. would like to thank the Observatoire de Besan\c{c}on for
supporting his work. The authors would also like to thank Leitherer et al.
for making their synthetic starburst models available to the community and
an anonymous referee for interesting comments and useful suggestions.
This research has made
use of the NASA/IPAC Extragalactic Database (NED) which is operated by the Jet 
Propulsion Laboratory, Caltech, under contract with the National Aeronautics 
and Space Administration.

\bsp

\label{lastpage}

\end{document}